\documentclass[aps,pra,twocolumn,showpacs,superscriptaddress,groupedaddress,a4paper,longbibliography]{revtex4-1}
\usepackage{amsmath}
\usepackage{epsfig}
\usepackage{graphicx}
\usepackage{color,float}
\usepackage{amssymb}
\usepackage{epstopdf}
\usepackage[pagewise,left]{lineno}
\usepackage[utf8]{inputenc}
\usepackage[T1]{fontenc}
\usepackage[colorlinks=True, linkcolor=blue, citecolor=blue]{hyperref}
\usepackage{xcolor}
\begin{document}
\title{Breakup of rotating asymmetric quartic-quadratic trapped condensates}
\author{Leonardo Brito$^{1}$}\thanks{brito@if.usp.br}
\author{Alex Andriati$^{1}$}\thanks{andriati@if.usp.br}
\author{Lauro Tomio$^{2}$}\thanks{lauro.tomio@unesp.br}
\author{Arnaldo Gammal$^1$}\thanks{gammal@if.usp.br}
\affiliation{$^{1}$Instituto de F\'{i}sica, Universidade de S\~{a}o Paulo, 05508-090 S\~{a}o Paulo, Brazil.\\
$^{2}$Instituto de F\'isica Te\'orica, Universidade Estadual Paulista, 01156-970 S\~ao Paulo, SP, Brazil.}
\date{\today}
\begin{abstract}
The threshold conditions for a rotating pancakelike asymmetric quartic-quadratic confined condensate to break in two 
localized fragments, as well as to produce giant vortex at the center within the vortex-pattern distributions, are investigated 
within the Thomas-Fermi (TF) approximation and full-numerical solution of the corresponding Gross-Pitaevskii (GP) 
formalism.  By comparing the TF predictions with the GP solutions,  in our investigation with two different quartic-quadratic 
trap geometries, of particular relevance is to observe that the TF approach is not only very useful to display the averaged 
density distribution, but also quite realistic in establishing the critical rotational conditions for the breakup occurrence and 
possible giant-vortex formation. It provides almost exact results to define the contour of the condensate distribution, even 
for high rotating system, after the system split in two (still confined) clouds. The applicability of the Feynman rule to the 
vortex distribution (full-numerical GP solutions) is also being confirmed for these non-homogeneous asymmetric trap 
configurations.  This study is expected to be relevant for manipulating the rotation and trap parameters in addition to 
Feshbach resonance techniques. It can also be helpful to define initial conditions for any further studies on dynamical 
evolution of vortex pattern distributions.
\end{abstract} 
\maketitle
\date{\today}

\section{Introduction}
The fragmentation phenomenon of objects due to their fast rotation is ubiquitous in physical systems, ranging from classical 
astronomical objects, such as galaxies and supermassive rotating stars~\cite{Maeder2009,2017Uchida,2017Chiappini}, to the 
microscopic world of quantum phenomena, in nuclear physics~\cite{1979Bohr,1998Clark}. In atomic-molecular physics,
this possibility could also be visualized by considering the experimental realization in Ref.~\cite{2020Guo} of  superfluid flow 
of quantum gas rotating in an anharmonic potential.  Theoretically, the occurrence of breakup in a rotating dipolar condensate 
was suggested in Ref.~\cite{Kumar2016}, when studying three-dimensional (3D) vortex structures in a rotating dipolar 
Bose-Einstein condensate (BEC).  For coupled condensed systems, the breakup can be easily understood as related to the miscibility 
properties of the mixture, such that the system split in two spatially separated condensates when going to the immiscible regime, 
in which the rotation can help to reduce the confinement of the two species  together~\cite{1998Barankov,2019Wilson,2012Web,2014Pattinson,2014Filatrella,2017Kumar,2019Kumar}.  
With single species atomic condensates, there are plenty of studies with different kind of rotating trap confinements, which allows 
the generation of multiple quantized vortices,  when considering enough strong power-law and anharmonic potentials~\cite{Fetter2001,2002Lundh,2003KavoulakisNJP,Fetter2005,2009Fetter}.  Among them, of particular interest are the 
investigations with the more experimentally feasible quartic-quadratic trap combinations~\cite{2002Lundh,2003KavoulakisNJP},
such that one can identify formations of circular super-flow and stable giant vortex, when considering fast and overcritical 
rotations~\cite{2001Recati,2002Kasamatsu,2003Engels,2003Fischer,2004Bretin,2004Aftalion,2004Simula,2004Baym,2011Correggi,2013Correggi,2019Adhikari}. 
However, in these cases, without interspecies interaction to play, we are not aware of specific approaches concerning the 
possibilities to fragment a condensate by introducing high rotation frequencies in a controlled way, with the fragments still kept 
confined.  In this direction, we can foresee the experimental possibilities, which are accessible, as one can follow from the recent 
studies performed with high rotation superfluids and condensates, generating long-lived dynamical rings~\cite{2020Guo,2019Pandey}. 
In order to generate long-lived stable vortices, the experiments are usually performed with repulsive two-body interactions, as being 
already well-known that vortices are generally unstable in rotating BEC with attractive 
interactions~\cite{2004Lundh,2006Carr,2008Sakaguchi,2015Sudharsan}, which are closely related to the condensate requirements for 
stability~\cite{1999Dalfovo,2009Fetter}.  

In the present work, we consider a BEC with repulsive two-body interactions, as our purpose is to study more 
closely the possibilities to break a stable condensate by increasing the rotation. For that an originally simple 
pancakelike two-dimensional (2D) trap confinement is assumed, which is obtained by a much stronger confinement 
in one of the directions, which can be easily implemented experimentally.  Our aim with such study is to fill a gap in the basic 
literature concerning the properties of rotating BECs~\cite{1999Dalfovo,2006Aftalion,2008Pethick,2009Fetter,2016Pitaevskii},
which can be relevant to control and calibrate possible experimental realizations when considering more involved 
studies on properties of ultra-cold condensed systems. By introducing such a breakup mechanism in an experimental 
apparatus, with the fragments of the condensate still being kept confined, one could have access to another 
mechanism to control the condensed cloud, in addition to the Feshbach resonance techniques~\cite{1998inouye,2010chin}. 
In view of the above motivations for considering a condensate breakup, we introduce a theoretical demonstration that such 
procedure can be easily implemented in a single non-dipolar condensate, induced by the rotation in an asymmetric trap. 
For that, we rely in the Thomas-Fermi (TF) approximation, for a 2D confined system, supported by the corresponding 
full-numerical results obtained via the usual Gross-Pitaevskii (GP) formalism.  
The numerical results are performed by imaginary time propagation approach~\cite{2006-Brtka} and, complementary, 
refined by fixed point methods~\cite{2009-Yang}.

In order to simplify our analysis and the system be experimentally feasible, 
we are assuming a strongly confined quasi-2D trap, considering two distinct 
quartic-quadratic non-symmetric shapes in the pancakelike perpendicular plane.
Both trap non-symmetric geometries are equally simple to be implemented in actual cold-atom experiments,
which are slightly different from the more symmetric quartic-quadratic confinements being so far 
investigated~\cite{2003KavoulakisNJP,2009Fetter,Fetter2005,2005Cozzini}. 
In both the cases that we are studying, a general double-well structure is assumed for the asymmetric confinement,
such that just by manipulating the rotation frequency parameter $\Omega$, one can deform and split the condensed 
cloud in two fragments, which can still be kept strongly confined within fixed two-well positions, which are affected   
by the $\Omega$ value. 
Of particular relevance for possible experimental proposals, as it will be further discussed, 
is the outcome of our study on the reliability of the TF approximation, which is quite robust and 
useful to define the limiting rotational conditions, even for the asymmetric trap structures under analysis, 
before and after the condensate splits in two fragments. 

By full-numerical solution of the GP formalism, we are also investigating the vortex structure patterns, which occur
in the two kind of asymmetric quartic-quadratic trap configurations being considered.  
The applicability of the well-known Feynman rule (FR)~\cite{1955-Feynman} in these quite asymmetric geometries
is well established, by verifying the vortex density numbers inside the confined condensed cloud, even when applying 
the rule to each fragment of the condensate, once it is appropriately defined regions with enough vorticity. The role 
played by hidden vortices (located in low-density regions) is also being characterized in our analysis.

In the next Sect.~\ref{sec2}, we present some details on the BEC formalism, with quartic-quadratic 
confinement, in which we are paying particular attention to the related TF description. 
All the numerical results for the GP formalism, as well as for the TF approach, are presented and 
discussed along this section, following the formalism.
In the Sect.~\ref{sec3}, we present our analysis on the vortex-pattern distributions verified in the
densities, in which we discuss the validity of the well-known Feynman rule for asymmetric shaped 
BEC confinement. Finally, in Sec.~\ref{sec4} we resume the work with our conclusions and perspectives for
further related studies.

\section{Quartic-quadratic trapped BEC}\label{sec2}
In the next, we present the usual basic quasi-2D GP formalism, obtained from a 3D reduction to a strongly 
pancakelike confined system in the $x_3-$direction. In the formalism two possible trap interactions are being 
considered in the $(x_1,x_2)-$plane. The breakup conditions are investigated in rotating condensate systems, 
in which we have verified that the usual TF approximation can successfully describe the density distribution, as 
well as it is quite helpful in defining the contours of the condensate, before and after the breakup.  Therefore, 
this approximation will be used in order to obtain the necessary rotating frequency in which the fragmentation 
of the condensate occurs in two parts. 

For a condensed system of particles having mass $m$, confined by a trap potential $ V_{3D}(x_1,x_2,x_3)\equiv 
 V_{2D}(x_1,x_2)+m\omega_3^2 x_3^2/2$,  where $\omega_3$ is the trap frequency along the $x_3-$direction, within a 
 rotating frame with angular frequency $\overline{\Omega}$ about $x_3$, where the the angular momentum operator 
 is $L_3={\rm i}\hbar\left(x_2\partial/\partial x_1 - x_1\partial/\partial x_2\right)$, the 2D reduction of the original 3D 
 stationary solution of the modified Gross-Pitaevskii (GP) formalism 
can be written, for the chemical potential $\mu_{2}\equiv\mu_{3}-\hbar\omega_3/2$, as
{\small\begin{eqnarray}\label{eq01}
\mu_{2}\psi_{2D}&=&\left[-\frac{\hbar^2}{2m}\nabla^2_{2D}+V_{2D}+  
\overline{\Omega} L_3 + g_{2D}\left|\psi_{2D}\right|^2
\right] \psi_{2D} ,
\end{eqnarray}
}where 
$\nabla_{2D}^2\equiv \frac{\partial^2}{\partial x_1^2}+\frac{\partial^2}{\partial x_2^2}$ and 
$\psi_{2D}\equiv\psi_{2D}(x_1,x_2)$ is the 2D wave function normalized to the number of atoms $N$
(the corresponding density given by $n_{2D}\equiv |\psi_{2D}|^2$), 
with $V_{2D}\equiv V_{2D}(x_1,x_2)$ being the 2D trap potential. 
 The strength of the nonlinear interaction, $g_{2D}$, related to the corresponding $g_{3D}$ 
due to the two-body $s-$wave interaction $a_s$, is given by
\begin{equation}\label{eq02}
g_{2D}\equiv  g_{3D} \sqrt{\frac{m\omega_3}{2\pi\hbar}}\equiv 
 \left(\frac{4\pi\hbar^2 a_s}{m}\right) \sqrt{\frac{m\omega_3}{2\pi\hbar}} .
\end{equation}
The total energy, related to (\ref{eq01}) can be written in terms of the 2D vector ${\bf r}$ 
and its conjugate momentum  
$\mathbf{p}$, as
{\small\begin{eqnarray}\label{eq03}
    E[\psi_{2D}] &=& \int \!\! \mathrm{d}^2 \mathbf{r} \left\{ \frac{1}{2m} \psi_{2D}^{*} 
    (\mathbf{p} - \mathbf{\overline{\Omega}} \times m\mathbf{r})^2 \psi_{2D}\right. \\
    &+&\left. \left[ V_{2D}- \frac{m\overline{\Omega}^2 r^2}{2} \right] |\psi_{2D}|^2 +
    \frac{g_{2D}}{2} |\psi_{2D}|^4 \right\}.\nonumber
\end{eqnarray}
}With the energy and length units given, respectively, by $\hbar\omega$ and $l_\omega\equiv \sqrt{\frac{\hbar}{m\omega}}$, 
we have $({x},{y})=(x_1/l_\omega,x_2/l_\omega)$, the GP formalism can be written in dimensionless 
units.  For that, we redefine the respective quantities, appropriately, as  
$\mu\equiv\mu_{2}/(\hbar\omega)$, $V\equiv V(x,y)\equiv V_{2D}/(\hbar\omega)$,  $\psi\equiv\psi(x,y)\equiv
l_\omega\psi_{2D}/\sqrt{N}$,
$\Omega\equiv\overline\Omega/\omega$ and $L_z=L_3/\hbar$.  
Therefore, from  Eqs.~(\ref{eq01})-(\ref{eq03}), we obtain the corresponding dimensionless equations, as
{\small\begin{eqnarray}\label{eq04}
\mu\psi=
\left[-\frac{1}{2}\left(\frac{\partial^2}{\partial {{x}^2}}+\frac{\partial^2}{\partial {{y}^2}}\right)+V
+\Omega L_z + g\left|\psi\right|^2\right]\psi,
\end{eqnarray}
\noindent }where $\psi$ is normalized to one. 
In the above, within our units, $\omega$ is an arbitrary frequency parameter, which is assumed to be perpendicular to
the $z-$direction, with $\lambda\equiv \omega_3/\omega$ being the trap aspect ratio. 
In this case, the dimensionless nonlinear strength carries the information about the
number of atoms $N$, trap aspect ratio $\lambda\equiv{\omega_3/\omega}$ and two-body 
$s-$wave scattering length $a_s$:
\begin{equation} \label{eq05}
g=N\sqrt{8\pi\lambda} \left(\frac{a_s}{l_\omega}\right).
\end{equation} 
We assume repulsive two-body interaction, with $a_s>0$ throughout this study, due to stability 
requirements of the condensate cloud, such that $g$ is always positive.
The dimensionless energy, corresponding to \eqref{eq03} is given by 
\begin{eqnarray}\label{eq06}
    {\cal E}[\psi] &=& 
    \int \!\! dx\,dy  \left\{ \frac{1}{2} 
    \left[\left|{\rm i}\frac{\partial\psi}{\partial x}-\Omega y\psi\right|^2 +
    \left|{\rm i}\frac{\partial\psi}{\partial y}+\Omega x\psi\right|^2\right]
    \right. \nonumber\\
    &+&\left. \left[ V(x,y) - \frac{{\Omega}^2 (x^2+y^2)}{2} \right] |\psi|^2 +
    \frac{g}{2} |\psi|^4 \right\},
\end{eqnarray}
from where we can observe that   
the trap potential is going to be effectively reduced as we increase the rotation frequency. 
About our full-dimension trap interaction, $V_{3D}(x_1,x_2,x_3)$, as usual we are assuming that the 
trap aspect ratio $\lambda=\omega_3/\omega$ is large enough in order to validate our quasi-2D geometry 
approach, in consonance with possible experimental realizations. For that, one can also realize that the
number of particles $N$ together with $\lambda$ and the two-body $s-$wave contact interaction $a_s$
can be conveniently adapted according to the system being investigated, by adjusting the strength $g$
[Eq.~\eqref{eq05}].

Within our aim to study the possibility for the rotation 
frequency to distort up to the breakup of the condensate, let us assume  non-symmetric 2D trap potential in the 
$(x,y)$ plane, given by a quadratic-quartic interaction. For that, let us fix  an harmonic oscillator trap in 
the $y-$direction, implying $\omega_y=\omega$ in our dimensionless units.  
For the quartic term, which impose a more strong confinement, we are going to consider 
two possibilities: The first, just $x-$dependent, with the second more symmetric, exchanging the 
$x^4$ by $r^4$. We can write both trap models as given by
\begin{eqnarray}\label{eq07}
V_{\sigma}({x},{y})&=&
\frac{y^2}{2}+\frac{b^2}{4} \left[\left(x^2-\frac{a^2}{b^2}\right)^2  + \sigma\left(r^4-x^4\right)\right],
\end{eqnarray} 
where $\sigma=0$ ($x-$quartic) and  $\sigma=1$ ($r-$quartic) define the two kind of traps. In both cases,
$b^2/4$ is the strength of the quartic term. The minima for $V_\sigma(x,y)$ are given by $x\equiv\overline{x}=\pm a/b$ and
$y=0$, along $x$ and $y$ directions, respectively.

In the following, we provide a detailed Thomas-Fermi approximation analysis for the density distributions, which will  
show up to be quite realistic as compared with the numerical results, when used to define the threshold conditions.

\subsection{Thomas-Fermi approximation.}
From Eq.~(\ref{eq03}), we can derive the TF approximation, when the atomic repulsive interactions are dominating,
by expanding the condensate to a mean radius that exceeds the mean oscillator 
length $l_\omega$. In this case, the expansion reduces the radial gradient of the density, with the kinetic 
energy becoming negligible relative to the trap and interactions, such that 
{\small \begin{eqnarray}\label{eq08}
{\cal E}_{TF}[\psi] &\approx& \int \!\! dx\,dy  \left\{ 
\left[ V_\sigma(x,y) - \frac{{\Omega}^2 r^2}{2} \right] |\psi|^2 + \frac{g}{2} |\psi|^4 \right\}.
\end{eqnarray}
}By minimizing this energy with respect to $|\psi|^2$ at fixed normalization, we obtain the 
TF density, which is a function of $\Omega$ through $\mu_{TF}\equiv\mu_{TF}(\Omega)$. It is given by 
\begin{eqnarray}\label{eq09}
n(x,y;\Omega)&\equiv&|\psi|^2=\frac{\mu_{TF}}{g}- \frac{1}{g}\left[ V_\sigma(x,y) - \frac{{\Omega}^2 r^2}{2} 
\right].
\end{eqnarray} 
We emphasize here that besides the double-well structure provided by $V_{\sigma}$, the centrifugal part 
effectively enlarge the distance between the minima, as can be seen from the term inside square brackets 
in the above Eq.~\eqref{eq09}, where for a fixed $\Omega$, the density $n(x,y;\Omega)$ have two 
maxima located at $(x,y)=\left(\displaystyle\pm\sqrt{\frac{\Omega^2 + a^2}{b^2}},0\right)$. In this case, the density at 
the center of the cloud, given by
{\small\begin{equation} \label{eq10}
n_{0}\equiv n(0,0;\Omega)\equiv \frac{\mu_{TF}-V_\sigma(0,0)}{g}
=\frac{1}{g}\left[\mu_{TF}-\displaystyle\frac{a^4}{4b^2}\right],
\end{equation}
}shall decrease by increasing the rotation, due to centrifugal forces, with the distances between the two
density maxima increasing. Within this dynamical process, $n_0$ can be zero, with the creation of
 a giant vortex at the center, as well as induce a breakup of the condensate in two clouds, which will 
 depend on the specific trap potential configuration.
So, the conditions to occur the breakup, or to generate the giant vortex, are established when the
effective chemical potential $\mu_{TF}-a^4/(4b^2)$ becomes zero ($gn_0=0$). However, 
by increasing even more the rotation $\Omega$, the effective chemical potential becomes negative,
with $gn_0$ defined by Eq.~(\ref{eq10}) representing this effective chemical potential, and not just  
the density at the center.

For the $V_{\sigma}$ traps defined in Eq.~\eqref{eq07}, the centrifugal contribution can either generate a 
giant vortex or split the cloud for $\sigma=1$ ($r-$quartic) case, in contrast to the $\sigma=0$ ($x-$quartic) 
case, where the cloud can split as well, but no giant-vortex phase is generated (as it will be shown). 
Therefore, a more relevant criterion is to analyze the contour of the Thomas-Fermi density in polar coordinates, 
as a function $r(\theta)$, since the breakup can easily be identified with the absence of real solutions at 
$\theta=\pm\pi/2$ (in view of the maxima distribution along the $x-$direction).  In order to support this TF criterion
to define the critical rotation conditions for the condensate breakup (both kind of potentials), as well as for the 
giant-vortex formation (case of $r-$quartic), we rely on the close agreement between
the TF approximation with the numerical GP solution when defining the condensate contour.
The two trap models defined in Eq.~\eqref{eq07}, for $\sigma=0$ ($x-$quartic case) and $\sigma=1$ ($r-$quartic case), 
will be treated separately in the next subsections.

\subsection{Case of $x-$quartic confinement} 
The contour of the cloud condensate, obtained by solving $n(x,y;\Omega) = 0$, before and after the breakup, 
is similar to what is known in mathematics as the {\it Cassini Ovals}~\cite{cassini}. Just at the breakup limit, the 
corresponding {\it oval} contour curve can also be identified with the {\it Bernoulli Lemniscate}. 
In polar coordinates, with $(x,y)\equiv [x(\theta),y(\theta)]\equiv[r(\theta) \cos{\theta}, r(\theta) \sin{\theta}]$, 
by applying the contour equation $n(x_c,y_c;\Omega) = 0$, we obtain the angular dependence of 
$r_c\equiv r_c(\theta)$, which is given by
{\small
\begin{equation}\label{eq11}
4gn_0 = b^2 r_c^4\cos^4\theta - 
2r_c^2[(1+a^2)\cos^2\theta-(1-\Omega^2)] .
\end{equation}
}The limit $\theta \to \pi / 2$, given by \eqref{eq11}, 
$r_c^2(\pi / 2)=2gn_0/(1-\Omega^2)$, reduces to zero
in the breakup, given by Eq.~\eqref{eq10},
\begin{equation}\label{eq12}
    n_0(\Omega_s) = 0 \rightarrow \mu_{TF}(\Omega_s) = \frac{a^4}{4b^2},
\end{equation}
where $\Omega_s$ is the critical rotation frequency to break the condensate  in two
separated clouds. 
The above relation expresses the fact that the chemical potential,
which is a general function of the rotation parameter $\Omega$ given by Eq.~\eqref{eq09}, 
at the breakup (when the density reduces to zero) is given by $\Omega_s$.  Therefore,
the dependence of the breakup condition on the condensate density parameter $g$ 
can only become explicit once a functional relation becomes established 
between $\Omega_s$ and $g$, as it will be shown.

The more general solution of Eq.~\eqref{eq11}, for the  contour $r_c$, is given  by 
 \begin{eqnarray}\label{eq13}
b^2\,r_{c,\pm}^2 \cos^2\theta 
    &=&1+a^2-(1-\Omega^2)\sec^2\theta
    \\   &\pm&
    \sqrt{ \left[{1+a^2-(1-\Omega^2)\sec^2\theta }\right]^2 + {4b^2gn_0}}
     ,\nonumber
\end{eqnarray}
where $r_{c,+}(\theta)$ and $r_{c,-}(\theta)$ are the two possible solutions. 
For $n_0>0$, only one solution is possible, $r^2_{c,+}(\theta)$, giving us  
the  full TF contour for $0\le\theta <2\pi$.   
For $n_0<0$ (after the breakup), two possible solutions emerge, with $r_{c,-}(\theta)$ 
corresponding to inner  and $r_{c,+}(\theta)$ to outer contour points. 

Therefore, aside of $\Omega_s$ being the critical rotation for breakup, it is also the threshold 
minus sign solution for $\Omega > \Omega_s$.
Specifically at the critical rotation frequency $\Omega_s$, with $\mu_{TF}$ given by 
Eq.~\eqref{eq12}, we have the contour delimited by the $+$ sign solution as
\begin{eqnarray}\label{eq14}
r_{c,+}^2(\theta) 
&=&2\frac{\sec^4{\theta}}{b^2}\left[a^2+\Omega_s^2-(1+a^2)\sin^2\theta \right]\nonumber \\
&\times&\Theta\left(\frac{a^2+\Omega_s^2}{1+a^2}-\sin^2\theta\right),
\end{eqnarray}
where $\Theta(\xi)$ is the Heaviside step function (=1 for $\xi\ge 0$, =0 for $\xi<0$). The above implies 
that, at the critical breakup frequency, the radial contour is limited within a fixed interval.
By applying the normalization in Eq.~(\ref{eq09}), together with (\ref{eq14}),
after the radial integration,  for the given trap parameters $a$ and $b$, 
we obtain the relation between the critical splitting rotation 
$\Omega_s$ and the density nonlinear parameter $g$, which is given by
\begin{eqnarray}\label{eq15}
{g b^4}&=&\int^{\theta_m}_0 d\theta \cos^4\theta\, r_c^6(\theta) 
=\frac{32}{105} \sqrt{\frac{(a^2+\Omega_s^2)^7}{1-\Omega_s^2}},
\end{eqnarray} 
where $\theta_m \equiv \arcsin\left(\sqrt{(a^2+\Omega_s^2)/(1+a^2)}\right) \in [0,\pi/2]$ is 
fixed by the Heaviside step function in Eq.~\eqref{eq14}.
The parameter $g$, as defined in (\ref{eq05}), is determined by the number of atoms $N$, the two-body scattering length
and the pancake 2D asymmetry $\lambda$. Therefore,  given $g$ and the trap parameters $a$ and $b$, we obtain analytically 
the critical frequency $\Omega_s$ for the condensate to split.
Within our purpose to verify the effect of a rotation in breaking a condensate, we are obviously considering that the
confining asymmetric double-well shaped trap \eqref{eq07} have the parameter-ratio $a/b$ such that the two minima 
position are close enough in order to have a continuous distribution of the condensate when $\Omega=0$.

\begin{figure}[h]
    \centering
    \includegraphics[scale=0.45]{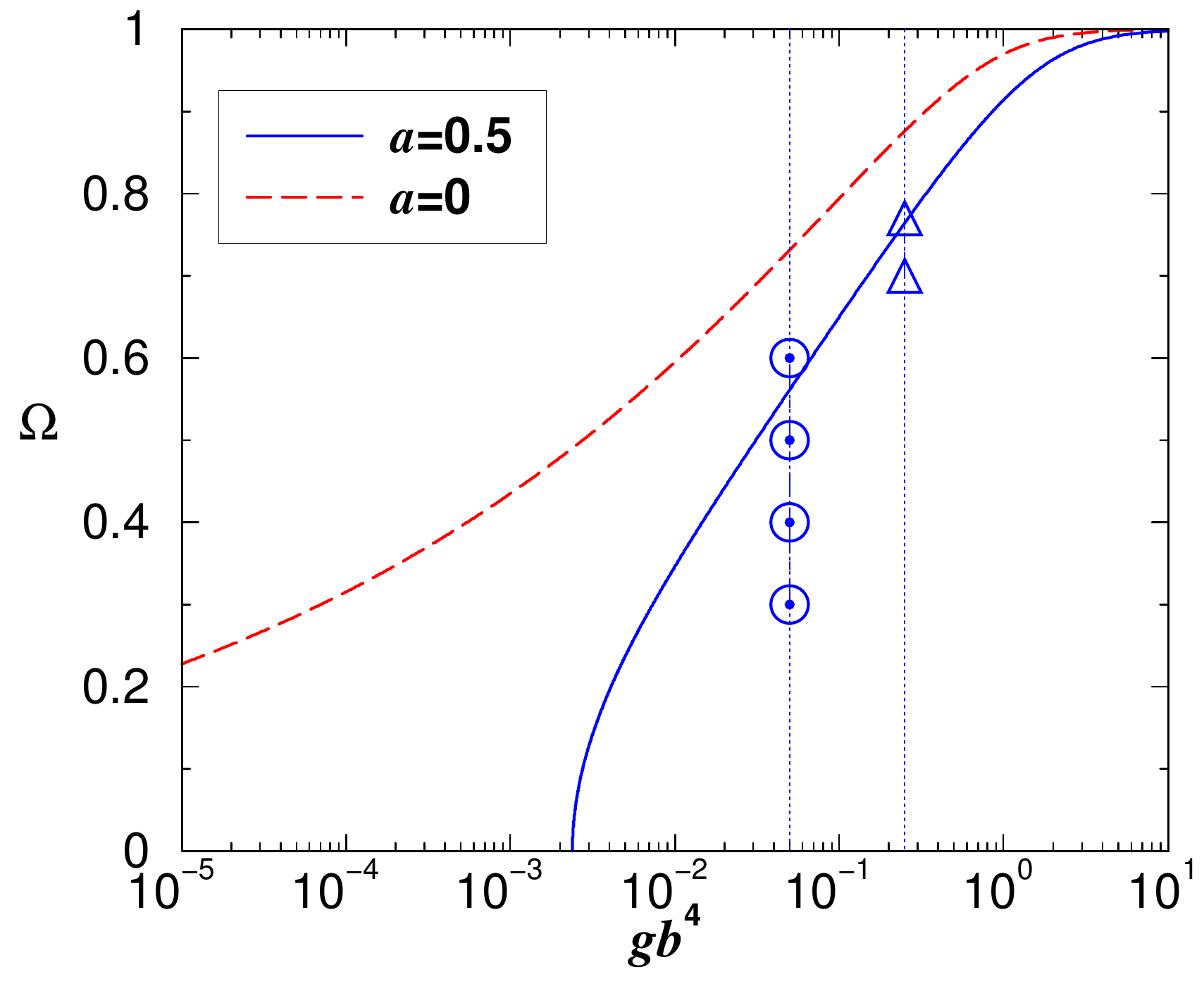}
    \caption{
    Rotation frequency $\Omega$ as function of  $gb^4$, for the trap defined by Eq.~\eqref{eq07}, 
    with  $\sigma=0$ and the nonlinear strength $g$ given by Eq.~\eqref{eq05}. 
    The curves, for $a=0$ (dashed) and $a=0.5$ (solid), refer to the TF lower limiting values, $\Omega_s$, 
    to split the condensate.  The circles (triangles) are indicating the positions of density-plot illustrations for $a=0.5$, shown 
    in 4 panels of Fig.~\ref{fig02} (3 panels of Fig.~\ref{fig03}), with $gb^4=0.05$ ($gb^4=0.25$),  where 
    $\Omega_s=$ 0.5616 (0.7640). With units defined in the text, all parameters are dimensionless.}
    \label{fig01}
\end{figure}
\begin{figure}[!htbp]
\centering
\includegraphics[scale=0.9]{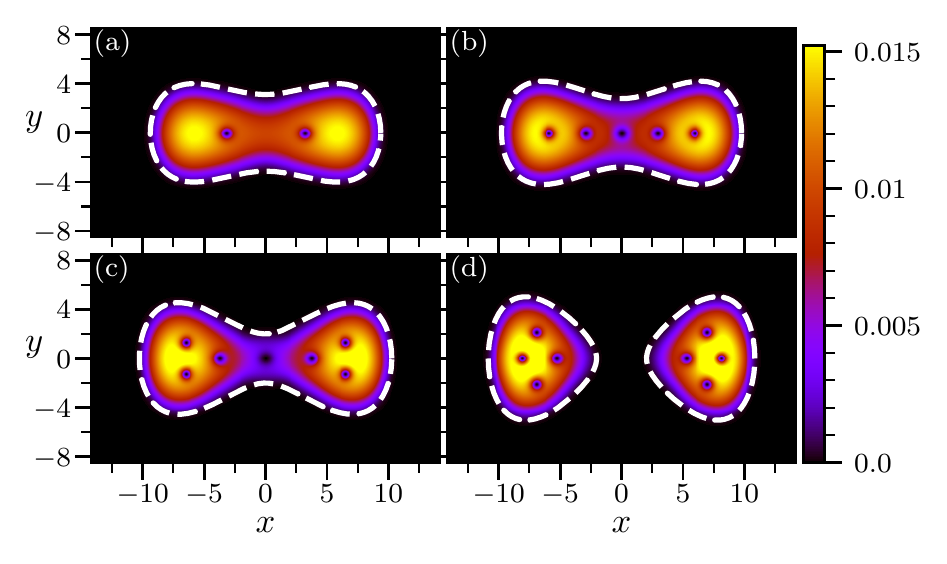}
\includegraphics[scale=0.44]{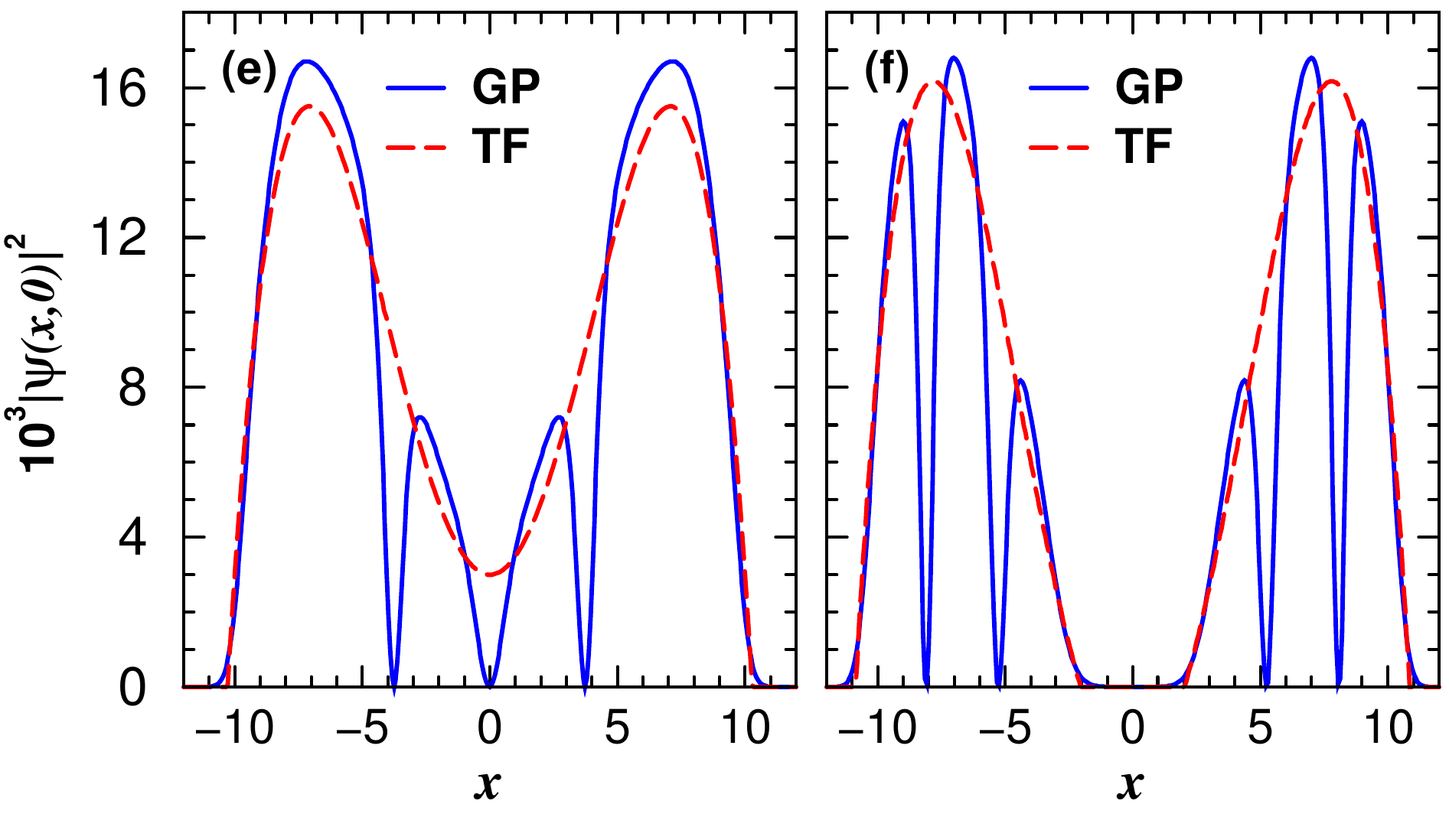}
\includegraphics[scale=0.12]{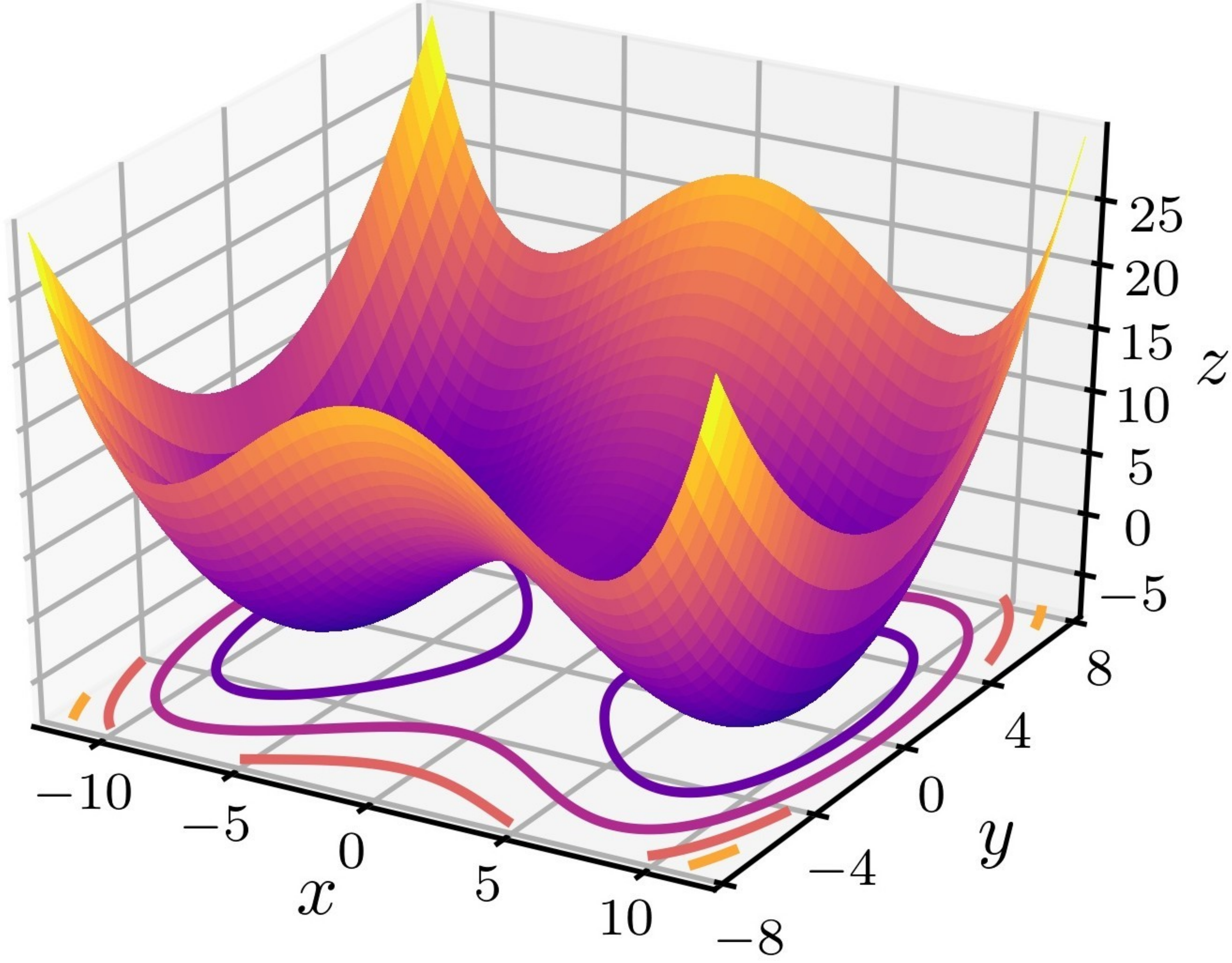} 
    \caption{
    The densities $|\psi(x,y)|^2$, obtained from the numerical GP solutions, are shown in the upper panels, for $\Omega = 0.3$ (a),
    $0.4$ (b), $0.5$ (c) and $0.6$ (d), with the TF approximation boundaries represented by white-dashed lines.
    In panels (e) and (f), corresponding to $y=0$ densities (c) and (d), respectively, we compare 
    TF (dashed curves) with GP (solid curves) results.  The other parameters are 
    $g = 500$, $b=0.1$ ($gb^4=0.05$) and $a = 0.5$.  The effective potential for the panel (d) is illustrated by the 
    surface plot [with $z\equiv V_0(x,y)-0.18 r^2$].  With the defined units, all 
    quantities are dimensionless.}
\label{fig02}
\end{figure}

\begin{figure}[!htbp]
    \centering   
\includegraphics[scale=.17]{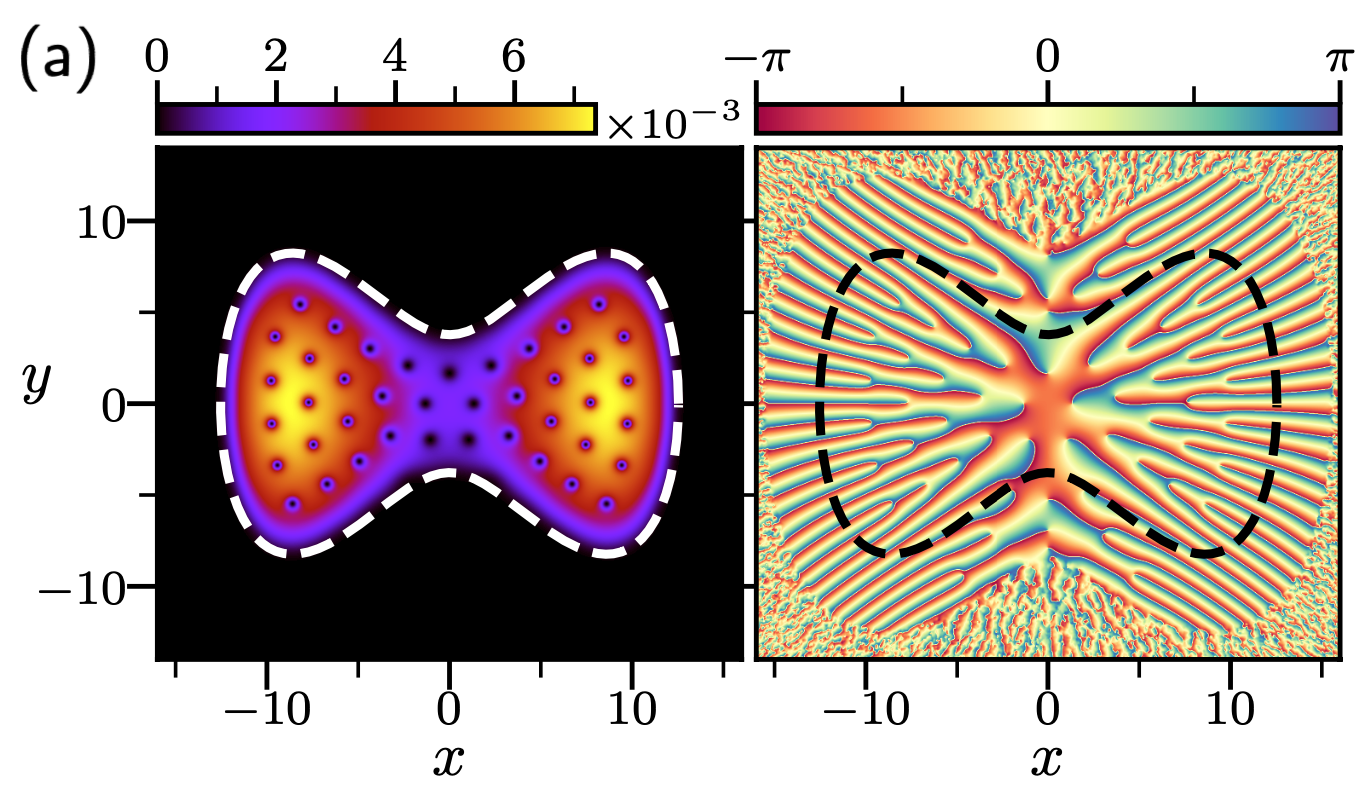}
\includegraphics[scale=.17]{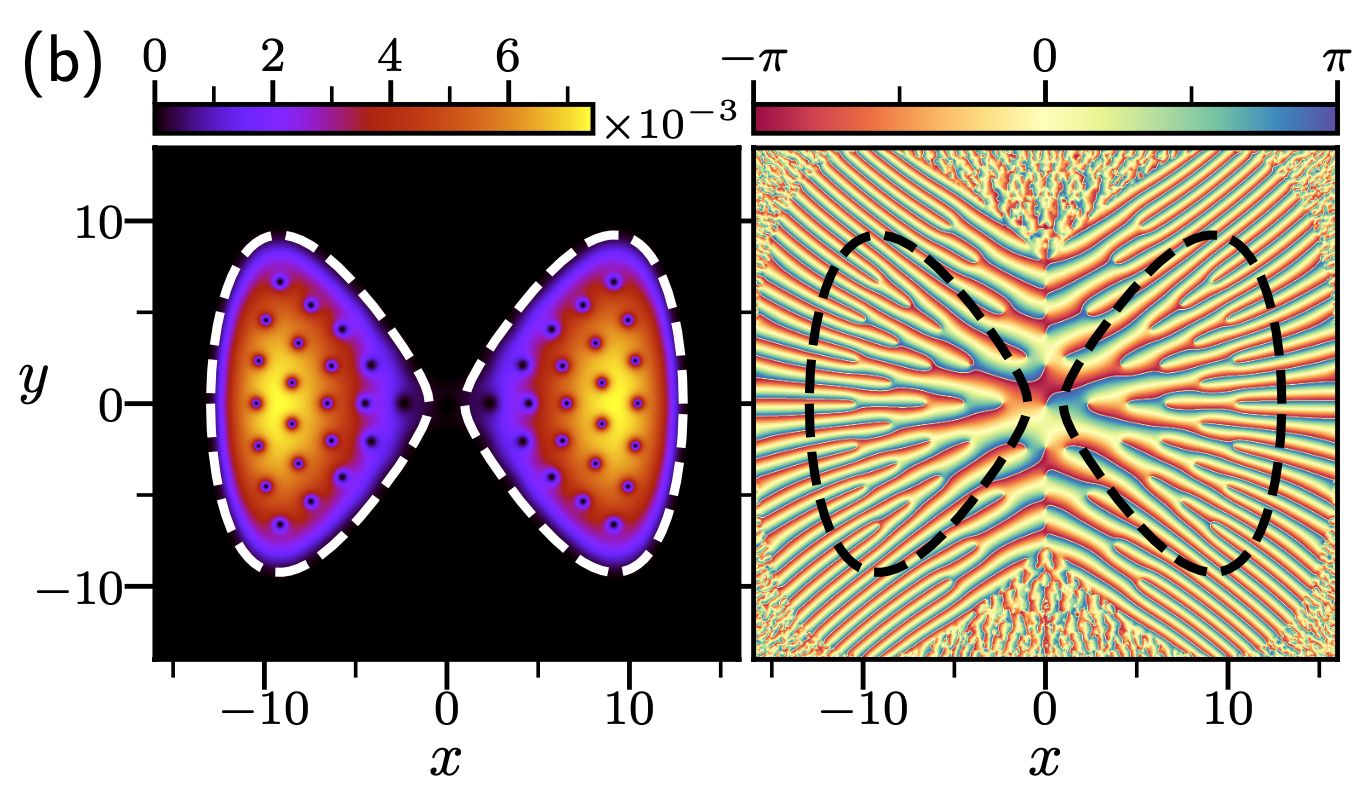}
    \caption{
Density distributions (left) are shown with the corresponding phase diagrams (right), for the nonlinear interaction $g =$ 2500 and rotation 
frequencies $\Omega =$ 0.7 [panel (a)] and 0.77 [panel (b)], whereas the breakup critical value is at $\Omega_s =$ 0.764. 
The trap parameters are the same as in Fig. 2. The TF boundaries are given by dashed lines,
with the upper color bars indicating the respective density and phase levels. With the defined units, all
quantities are dimensionless.}
\label{fig03}
\end{figure}

In Fig.~\ref{fig01} we are resuming the TF results for $\Omega_s$ as a function of $g b^4$, by considering two indicative 
fixed values of the parameter $a$ (0 and 0.5).  Few sample results for the densities, showing how the rotation is affecting the 
dynamics of the condensate, are also illustrated in Figs.~\ref{fig02} (with $g=500$) and \ref{fig03} (with $g=2500$), considering
numerical GP results.  In both the cases,
we have $a=0.5$ and $b=0.1$, such that the minima of the quartic interaction are at $x=\pm 5$.  The corresponding parameter 
positions of these results ($gb^4=$0.05 and 0.25, respectively) are also indicated inside the panel of Fig.~\ref{fig01}. 
The corresponding breakup rotational parameter are $\Omega_s=$ 0.5616 and 0.7640.

In Fig.~\ref{fig02}, we show the numerical GP simulations together with the TF boundaries for 
the densities, as considering four panels with different arbitrary values of $\Omega$, from 0.3 till 0.6. 
This figure illustrates how the breakup occurs in a condensate trapped by $V_0(x,y)$ given by Eq.~\eqref{eq07},  
induced by the rotation frequency, which is reducing the strength and shifting the position of the minima of the 
corresponding effective double-well potential. For a better quantitative analysis, we have also included the panels (e) and
(f) corresponding to $|\psi(x,0)|^2$ of panels (c) and (d). 
The effective potential related to panel (d) is also being represented in the bottom part of the figure by a surface 3D plot.
In particular, we consider remarkable the agreement between the numerical GP computed results with the corresponding TF
approximation estimates, when applied to the contour of the condensate (inside which we can find most of the density 
distributions).  
Such results obtained by using the TF approximation, as compared with the numerical computed ones, are indicating 
that it can be quite useful in order to verify the exact rotation frequency, $\Omega_s$, at which one should expect the 
condensate being fragmented in two pieces, once considered fixed the potential parameters.  
The results presented in the panels (e) and (f) are particularly significant to support the TF approximation in order 
to define the breakup limiting conditions. 
In spite of the obvious limitations in describing the vortex pattern structure,
the TF approximation provides an averaged good agreement with the numerical GP results for the internal averaged 
density distributions, remarkably reproducing the corresponding contour borders.

In Table~\ref{tab:Mu}, the quantitative deviations between the TF and the GP numerical results are verified for the chemical potentials
corresponding to the four specific cases shown in Fig.~\ref{fig02}. Among the results, we also add the ones 
corresponding to the total energy, together with the cases with $\Omega=0$. 
\begin{table}[ht]
\caption{
The chemical potentials (GP numerical simulations and TF approximation), for $\Omega=0$ and the frequencies considered 
in Fig.~\ref{fig02} are given in this table, together with the corresponding GP Energy values.
The units are $\hbar\omega$ for the energy and chemical potentials,  with $\omega$ for the frequency $\Omega$.}
\label{tab:Mu}
\centering
    \begin{tabular}{ l c c c }
    \hline\hline
$\Omega$&$\mu_{TF}$        &$\mu$ &   Energy      \\ \hline \hline
          0  & 7.332   & 7.395 & 4.849\\
        0.3  &  5.994  & 6.469  &  3.962  \\
        0.4  &  4.813  & 5.407  & 2.955  \\
        0.5  &  3.061  & 3.801  & 1.271  \\
        0.6  &  0.343  & 1.215  &  -1.490 \\
\hline\hline
    \end{tabular}
\end{table}

From the GP results shown in Fig.~\ref{fig02}, we observe the crescent number of  vortices formed inside 
the condensate as we increase the rotation. The linear increasing behavior of the vortices number with the frequency
 is clearly being broken near the breakup limit, giving us some indication of the occurrence of possible hidden 
vortices at the low-density region. This behavior will be further discussed in the next section, within our analysis of vortex pattern 
distributions.   

In order to  improve our illustrations of the $x-$quartic confining potential, with densities in which we
can verify more clearly how the vortex pattern distributions occur before and after the breakup,  we include the  
Fig.~\ref{fig03}, where we consider two specific cases with frequency rotations $\Omega=$ 0.70 (a) and 0.77 (b), 
with the nonlinear parameter, $g=2500$, which is much larger than in the case of Fig.~\ref{fig02}. However, in both the
cases, we are keeping the same trap parameters, $a=0.5$ and $b=0.1$, with the potential minima at $x=\pm 5$. 
These plots, in which a richer pattern structure can be observed, were obtained by the GP numerical simulations 
for the densities (left panels), with the corresponding phase diagrams (right panels). The boundary definitions, indicated
inside the panels, were obtained by using the TF approximation.
As shown in the panel (b), with rotation frequency $\Omega=0.77$,  the condensate is already broken in two  fragments. 
Due to the rotation $\Omega=0.77$, the positions of the minima of the effective potential move to $x=\pm 9.18$, 
with both fragments presenting about the same kind of triangle vortex pattern distributions. 
As for $\Omega=0.70$, panel (a), we observe about the same vortex pattern, with the minima of the effective 
potential being at $x=\pm 9$. 
In both phase diagrams we observe the existence of hidden vortices which are located in the low-density regions, 
along the $y-$ axis, close to $x=0$. More specifically, these hidden vortices are located outside the region defined by 
the TF approximation, as they are manifested only by solving the full GP formalism.

\begin{figure}[h]
    \centering
\includegraphics[scale=0.35]{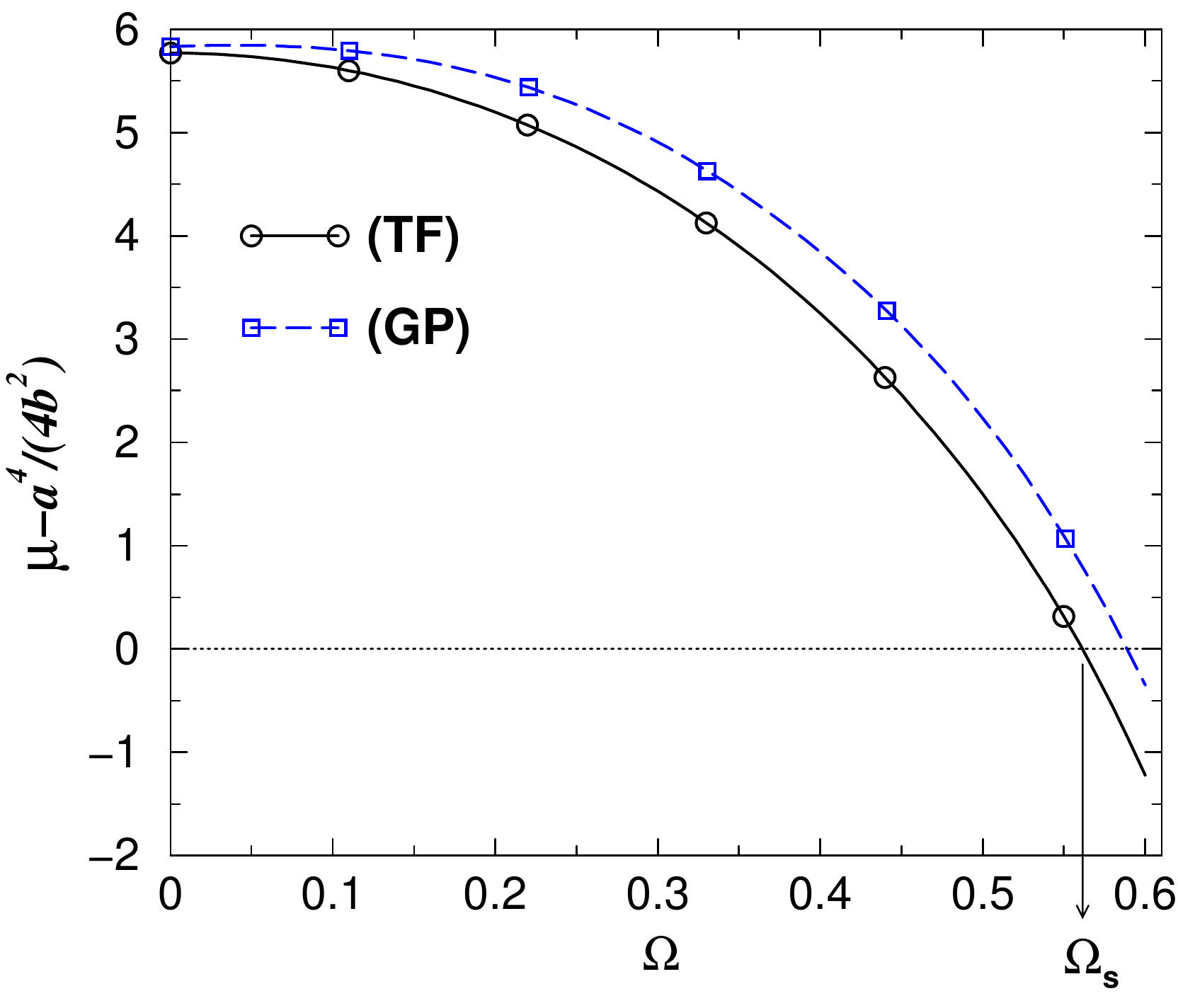}
    \caption{The dependence of the effective chemical potentials [GP and TF results], given by $\mu-{a^4}/{(4b^2)}$,
     on the rotation frequency $\Omega$ for the trap parameters $a=0.5$, $b=0.1$. 
    The  TF  (GP) results are indicated, respectively, by the circles (squares), being fitted by solid (dashed) lines.
    At $\Omega=0$, $\mu_{0}=$7.332 (TF) and 7.395 (GP).  For zero density, when $\mu=(5/4)^2$, we obtain  
    $\Omega_s=0.5616$, within the TF approximation; and 0.590 within the GP result. With the given units, all
   quantities are dimensionless. }
\label{fig04}
\end{figure}
\begin{figure}[t]
    \centering
    \includegraphics[scale=0.25]{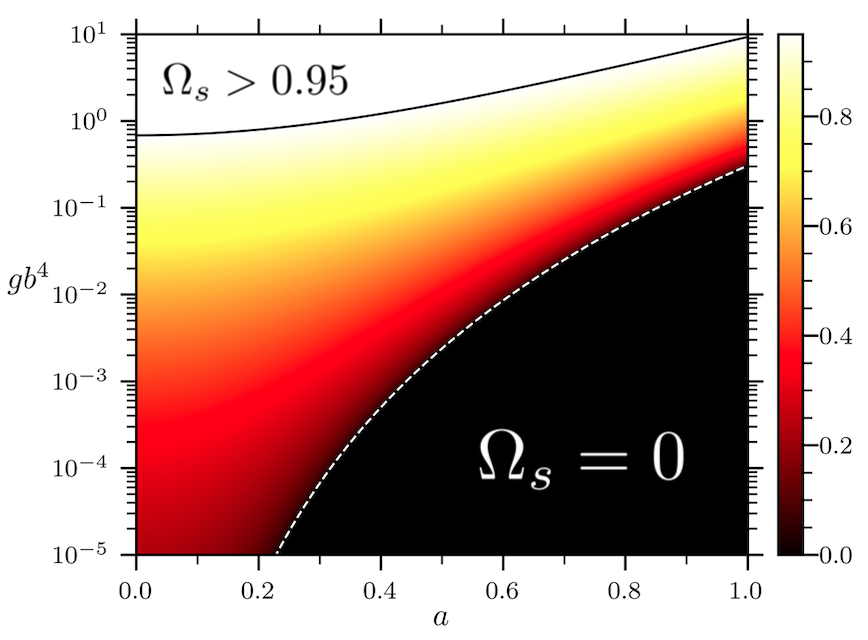}
    \caption{For the $x-$quartic trap potential [Eq.\eqref{eq07} with $\sigma=0$], we show by a color-level map 
    (indicating the necessary split frequency $\Omega_s$) the corresponding parametric region for which a rotation is 
    required for breaking the cloud. The upper white region corresponds to $\Omega_s > 0.95$ at which the condensate 
    is likely to fly apart in the $y-$direction before the split occurrence. Below the white-dashed line ($\Omega_s=0$), the
    cloud is already spatially separated by the two-well shaped trap. With given units, all quantities are dimensionless.}
    \label{fig05}
\end{figure}
By considering the same potential parameters used in Table~\ref{tab:Mu} and in Fig.~\ref{fig02}, with $a=0.5$ and $b=0.1$, 
we also present in Fig.~\ref{fig04} the corresponding behavior of the 
effective chemical potential, given by $\mu - {a^4}/{(4b^2)}$, in terms of the rotation frequency $\Omega$, 
which is zero at the condensate breakup, when $\Omega=\Omega_s=0.5616$.
For that we consider the TF approximation results together with numerical solutions obtained
from the GP equation. The results can be well fitted by approximate analytical polynomial expressions.
From these analytical results, we observe the quite small deviation of the TF critical breakup rotation limits, 
$\Delta\Omega_{s}=0.028$.
We also observe that the deviation of the TF chemical potential $\mu_{TF}$, from the GP values 
$\mu$ becomes larger for fast rotations, which one could expect due to the fact that, by increasing the rotation, 
more vortices are emerging in the condensate, with the effective confining interaction being diminished in comparison with 
the kinetic energy term. This is evidenced by the density plots shown in Fig.~\ref{fig02}.  

In Fig.~\ref{fig05}, a diagrammatic picture shows the relation between the trap parameters $gb^4$ and $a$ for 
the $x-$quartic potential [Eq.~\eqref{eq07} with $\sigma=0$], indicating the necessary rotation $\Omega_s$ 
to split the condensate.  $\Omega_s$ is represented by the color density, considering the 
upper-white region  $\Omega_s>0.95$ the region at which the condensate likely flies apart before the breakup. 
The parameter region at which the condensate is already broken by the potential parameters is shown below
the dashed line (where $\Omega_s=0$).

\subsection{Case of $r-$quartic confinement} 
By considering the confining potential given by Eq.~(\ref{eq07}) with $\sigma=1$, we have a trap with
quadratic+quartic radial symmetry in the $(x,y)$ plane, with the symmetry broken in the $x-$direction,
which can be written as
\begin{eqnarray}\label{eq16}
V_{1}(x,y)&=&
\frac{r^2}{2}+\frac{b^2}{4}r^4-(1+a^2)\frac{x^2}{2}  + \frac{a^4}{4b^2},
\end{eqnarray} 
Different from the previous $x-$quartic trap, in this case we have also a quartic component in the $y-$direction.  
The $|x|$ and $|y|$ extreme limits for non-zero density, obtained by using the Eq.~(\ref{eq09}) are defined by   
\begin{eqnarray}\label{eq17}
n(x,y)&\equiv&|\psi|^2=\frac{\mu_{TF}}{g}- \frac{1}{g}\left[ V(x,y) - \frac{{\Omega}^2 r^2}{2} 
\right] \\
&=& n_0-\frac{1}{2g}\left({{y}^2 -{a^2x^2}+\frac{b^2}{2} r^4 - \Omega^2 r^2}\right) ,
\nonumber,
\end{eqnarray} 
where $gn_0\equiv\left[\mu_{TF}-a^4/(4b^2)\right]$ as also given by Eq.~(\ref{eq10}).
Analogously as in the $x-$quartic case, we could follow by using polar coordinates.
However, before that, it may be useful to verify the extremes of non-zero density $n\equiv n(x,y)$.
So, by deriving in the $x-$coordinate, we can identify a minimum at $x=x_0=0$ with two maxima at 
$x=x_\pm$ of (\ref{eq17}), given by
\begin{eqnarray}\label{eq18}
x_\pm&=& \pm \sqrt{\frac{\Omega^2+a^2}{b^2}- y^2},\;\; x_0=0.
\end{eqnarray} 
Similarly, by deriving in the $y-$coordinate (given $x$), we can identify the existence of 
two maxima at $y=y_\pm$ with a minimum at $y=y_0=0$, only when $\Omega^2>1+b^2x^2$, 
\begin{eqnarray}\label{eq19}
y_\pm&=&\pm\sqrt{\frac{\Omega^2-1}{b^2} - x^2},\;\; y_0=0.
\end{eqnarray} 
Combining both results, we verify that for $\Omega^2>1$ we can have extrema (two maxima
and one minimum) only in the $x-$direction, located at $(x_\pm,0)$ ($\theta=0$), as the maxima obtained
by (\ref{eq19}) are located outside the positions where the previously broken condensate parts are located,
as we can see by replacing $x_\pm^2$ in (\ref{eq19}).

Now, let us consider the density contour given by Eq.~\eqref{eq17} in polar coordinates, 
$(x,y)\equiv [x(\theta),y(\theta)]$ $\equiv [r(\theta)\cos\theta,r(\theta)\sin(\theta)]$, such that
$r_{c}\equiv r_c(\theta)$ is given by 
\begin{equation} \label{eq20}
4 g n_0 = b^2 r_c^4 - 2 r_c^2 \Big[ (1+a^2)\cos^2{\theta}-(1-\Omega^2)\Big],
\end{equation}
{\small\begin{eqnarray}\label{eq21}
 b^2 r_{c}^2
    &=&  
{(1+a^2)\cos^2\theta-(1-\Omega^2)} 
    \\   &\pm&
    \sqrt{ \left[{(1+a^2)\cos^2\theta-(1-\Omega^2) }\right]^2 + {4b^2gn_0}
    } .\nonumber
\end{eqnarray}
}We should emphasize that the necessary and sufficient condition for the breakup is that no real solution
occurs for $r_c^2(\pm\pi/2) \equiv y_c^2$ (implying density zero along the $y$ axis, not only at $y=y_c$). 
In the $x-$quartic case ($\sigma=0$) the $n_0 = 0$ condition is enough for the 
breakup, with the frequency being limited to $\Omega\le 1$, as shown. 
Nevertheless, differently from $x-$quartic case, as we take $\theta = \pm\pi / 2$ in 
Eq.~\eqref{eq20} we still end up with a bi-quadratic equation for the radius, with the
Eq.~\eqref{eq21} simplified as
{\small
\begin{equation} \label{eq22}
b^2y_{c}^2 = {(\Omega^2-1)} \pm \sqrt{{(\Omega^2-1)^2} + {4b^2\left[\mu_{TF}-\left(\frac{a^4}{4b^2}\right)\right]}}.
\end{equation}
The existence of solutions for this equations implies that, at $\theta=\pm\pi/2$, the contour is 
connected. So, next we consider all the possibilities in (\ref{eq22}),
where $gn_0=\mu_{TF}-a^4/(4b^2)$:
If $gn_0 > 0$ only the $+$ sign solution can exist, corresponding to a connected contour in the TF description.
However, if $gn_0 \le 0$, the existence of real solutions (characterizing contour connected) 
depends on $\Omega$: For $\Omega < 1$, neither the $+$ nor the $-$ solutions are possible, characterizing 
the breakup of the cloud.  However, for $\Omega > 1$, both solutions are possible,  
if $0>4b^2g n_0 > -(\Omega^2-1)^2$,  corresponding to the upper and lower solutions of
the connected contour, which is identifying the giant-vortex formation at the center. 
The other possibility, when $4b^2g n_0<-(\Omega^2-1)^2$,
there is no real solution for Eq.~\eqref{eq22}, which will corresponds to the breakup. 
Summarizing:}
\begin{equation}\label{eq23}
    \begin{array}{c|l|l}
        \Omega > 1 & \displaystyle{-(\Omega^2-1)^2} < 4b^2gn_0 \leq 0 & \textrm{Giant vortex} \\
        & \displaystyle{-(\Omega^2-1)^2} \geq 4b^2 g n_0  & \textrm{Breakup} \\ \hline
        \Omega \leq 1 & gn_0(\Omega) \leq 0 & \textrm{Breakup}
    \end{array}
\end{equation}
The critical values for the rotation parameter $\Omega$ are corresponding to the transitions among the phases, which 
are obtained when the equality holds in the above relations, being defined as $\Omega_G$ for the Giant-vortex
formation (which can happen only for $\Omega>1$), and $\Omega_s$ the critical frequency to split the cloud.

Differently from the first trap studied, in the one defined by Eq.~\eqref{eq16}, the breakup regime can be achieved for larger values 
of $b$ since the condensate will no longer fly apart due to the presence of quartic confinement also in the $y$-direction.

In the above, we have considered the possible breakup of the condensate (as well as the giant-vortex formation in the origin)  
along the $y-$axis (considering $\theta=\pm \pi/2$, or $x=0$), when by increasing the rotation
(with $\Omega>1$) the quadratic confinement in the $y-$direction is reduced till the condensate split in two fragments. 
However, a similar situation could occur in the $x-$direction, if the asymmetry of the trap potential which is holding the
two fragments can allow minima for the density at $y=0$ $(\theta=0)$ for fixed values of $\pm x$, in Eq.~\eqref{eq21}.
This should be given by the following second-order equation:
{\small
\begin{equation} \label{eq24}
b^2x_{c}^2 = {(\Omega^2+a^2)} \pm \sqrt{{(\Omega^2+a^2)^2} + {4b^2gn_0(\Omega)}} .
\end{equation}
} For larger values of $\Omega^2$ ($\gg a^2$), the Eqs.~\eqref{eq22} and 
\eqref{eq24} provide the same results for $y_c$ and $x_c$, respectively. 

In the following, let us first consider $n_0(\Omega)=0$ in Eq.~\eqref{eq21}, which give us 
a critical $\Omega_c$. This is the split critical frequency $\Omega_s$ for $\Omega\le 1$. However, 
for $\Omega\ge 1$ it is a critical rotation for a giant vortex to appear, such that $\Omega=\Omega_{G}$.
For $n_0(\Omega_c)=0$, the critical condition gives
 {\small \begin{eqnarray}\label{eq25}
    r_{c}^2(\theta)
    &=&\frac{2(1+a^2)}{b^2} \left[C(\Omega_c)
-\sin^2\theta\right]
    \Theta\left(C-\sin^2\theta\right),        \nonumber\\
    &&{\rm with}\;\;\; C\equiv C(\Omega_c)\equiv \frac{a^2+\Omega_c^2}{1+a^2} .
\end{eqnarray}
}In the above, the Heaviside step function is only effective in case $\Omega_c\le 1$, when
$C$ can define an upper limit for the angle $\theta$, given by $\theta_m$.
In such a case, we can assume $C\equiv \sin^2\theta_m$ for $\Omega_c\le 1$.
In the other case, for $\Omega_c\ge 1$, we keep $C$ as defined in Eq.~\eqref{eq25}.

By applying the normalization condition in Eq.~\eqref{eq17}, together with 
Eqs.~\eqref{eq20} and \eqref{eq25}, we obtain
{\small \begin{eqnarray}\label{eq26}
gb^4&=& \frac{b^6}{24} \int^{\pi}_0d\theta r_{c}^6\nonumber\\
&=&\Theta(1-\Omega) \frac{2\left(1+a^2\right)^3}{3} \int^{\theta_m}_0d\theta
 \left[C-\sin^2\theta\right]^3 +\nonumber\\
&+&\Theta(\Omega-1) \frac{2\left(1+a^2\right)^3}{3} \int^{\pi/2}_0d\theta
 \left[C-\sin^2\theta\right]^3\\
 &=& \frac{\left(1+a^2\right)^3}{48}
 \left\{ -(2\theta_m) \cos(2\theta_m) 
  \left[2\cos^2(2\theta_m)+{3}\right]
   \right. \nonumber\\
  &+&
  \left.\left[{ \left({5}-\frac{11}{3}\sin^2(2\theta_m)\right){\sin(2\theta_m)} 
  }\right]\right\}\Theta(1-\Omega)\nonumber\\
    &+&
 \frac{\pi\left(1+a^2\right)^3}{3} \left\{\left(C-\frac{1}{2}\right)
 \left[C^2-C+\frac{5}{8}\right]\right\}
\Theta(\Omega-1)\nonumber
 \end{eqnarray}
}In the first case ($\Omega\le 1$), when the integral is limited by $\theta_m$,  we can identify
$\Omega_c$ with the critical splitting frequency $\Omega_s$. 
And, in the second case, for $\Omega\ge 1$, the upper limit of the integral is fixed at $\pi/2$, with the
critical value corresponding to the creation of a {\it giant vortex}, such that we identify $\Omega_c$
with $\Omega_G$, which is within the definition of $C$, given in Eq.~\eqref{eq25}. Therefore, in 
the above we have $C\equiv{(a^2+\Omega_G^2)}/{(1+a^2)}$.

Some specific cases: \\
\noindent (i) With $\Omega^2\le 1$, we notice that in the non-rotating case  
$\Omega=0$, for $a^2=1$, we obtain $gb^4= {2}/{9}$.  

\noindent (ii) With $\Omega\ge 1$,  for the giant-vortex critical frequency,
when $\Omega_G^2=1$,  we obtain $gb^4={5\pi(1+a^2)^3}/{48}$  (= 0.639, if $a=1/2$);
and with $\Omega_G = 2$, $gb^4=52.11$ (with $a=1/2$) and 46.27 (with $a=0$).

Next, we consider the breakup for the second case, $\Omega^2\ge 1$, which is obtained by increasing the rotation 
frequency beyond $\Omega_G$,  till the condensate split in two fragments. 
So, the density at $x=0$ (for any value of $y$), which is given by  
{\small \begin{eqnarray}\label{eq27}
n(0,y)&\equiv&\frac{1}{g}
\left[\left(\mu_{TF}-\frac{a^4}{4b^2}\right)-{\frac{b^2}{4} y^4 -\frac{1- \Omega^2}{2} y^2}
\right],\end{eqnarray}
}becomes zero along the $y-$axis, at a critical frequency $\Omega=\Omega_s>\Omega_G$, when 
$g n(0,y)=0$.  This condition is expressed by $4b^2gn_0=-(\Omega_s^2-1)^2$ in \eqref{eq23},
implying in $b^2y_c^2 = (\Omega_s^2-1)$ at the critical breakup limit.
Next, by following the same procedure as before, we consider the corresponding contour $r_c(\theta)$.
By applying the normalization (\ref{eq17}), 
we obtain
{\small\begin{eqnarray}\label{eq28}
gb^4&\equiv&\frac{2}{3}(1+a^2)^3 \int_0^{\pi/2}d\theta 
\cos^3\theta\sqrt{\left[\frac{2\Omega^2+a^2-1}{a^2+1}-\sin^2\theta\right]^{3}}.
\nonumber\\
&=&\frac{2(1+a^2)^3}{3} 
\int_0^{1}dz(1-z^2){\left[\zeta^2+1-z^2\right]^{3/2}}\big|_{\zeta^2\equiv{\frac{2(\Omega^2-1)}{a^2+1}}}
\nonumber\\
&=&\frac{(1+a^2)^3}{72 }\left[\zeta\left(3\zeta^4+22\zeta^2+15\right)
\right.\nonumber\\&-&\left. 3(\zeta^2-5)(\zeta^2+1)^2 \cot^{-1}\zeta\right].
\end{eqnarray} 
}Therefore, in case $\Omega^2=1$ we have $\zeta=0$, with $gb^4={5\pi}{(1+a^2)^3}/{48}, $ in agreement with
the previous result we have obtained for $\Omega^2\le 1$.

\begin{figure}[h]
    \centering
    \includegraphics[scale=0.45]{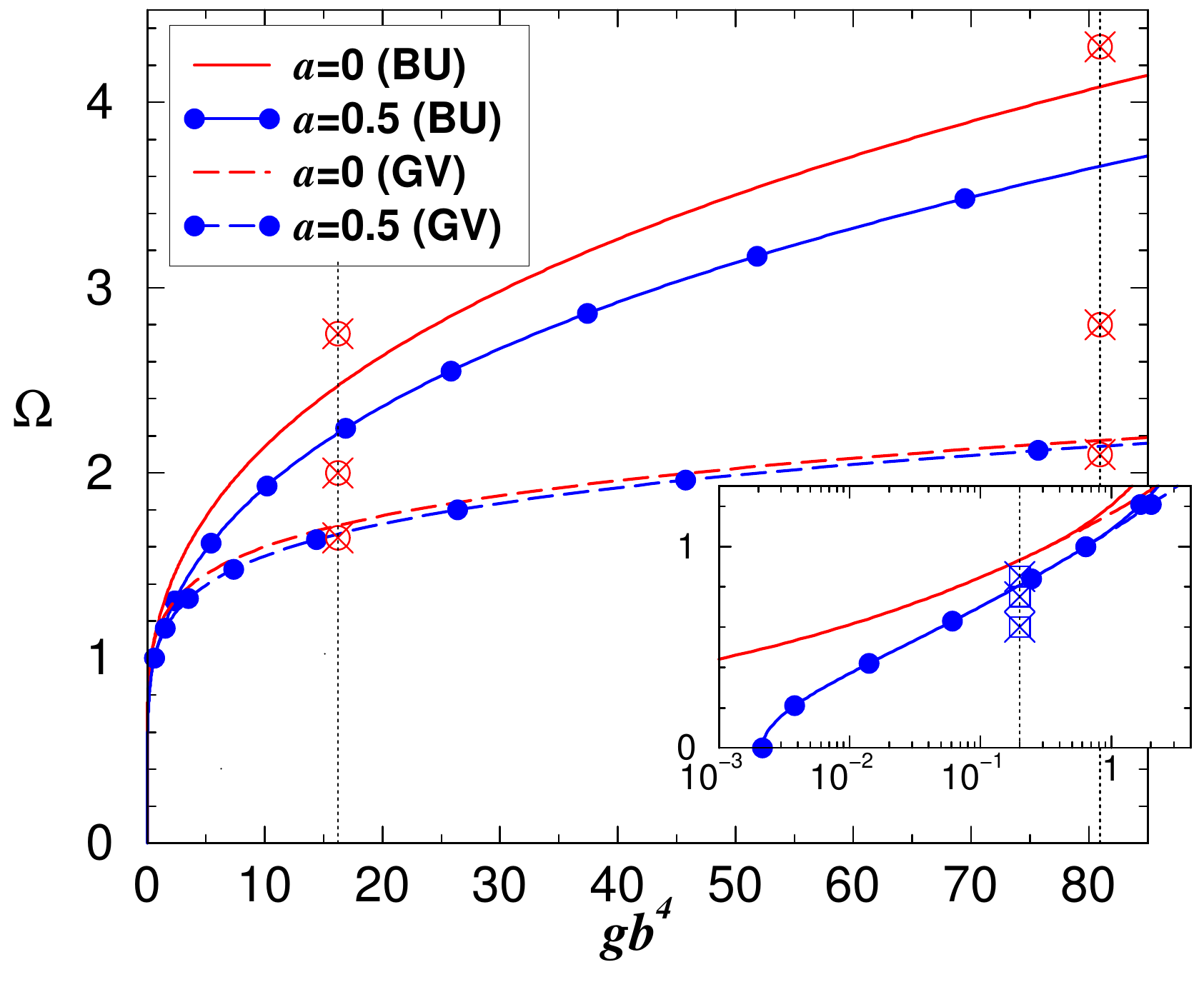}
    \caption{The condensate critical frequencies $\Omega$ to originate the giant vortex ($\Omega_G$) (GV, dashed lines) 
    and for the breakup ($\Omega_s$) (BU, solid lines) for $a=0$ and $a=0.5$ (line with bullets).  The crossed-circles are indicating 
    the coordinate positions of density plots shown in the three panels of Fig.~\ref{fig09} and also for the density plots given in the 
    three lower panels of Fig.~\ref{fig08} (indicated by the dotted lines at $gb^4=$16.2 and 81, respectively). 
    The inset (in log scale) shows the corresponding behavior for lower $\Omega<1$, with the crossed-squares indicating the coordinate 
    positions of the density plots shown in the upper panels of Fig.~\ref{fig08} (indicated by the dotted line at $gb^4=0.2$). 
    With the defined units, all quantities are dimensionless.
} \label{fig06}
\end{figure}

\begin{figure}[!htbp]
    \centering
    \includegraphics[scale=.25]{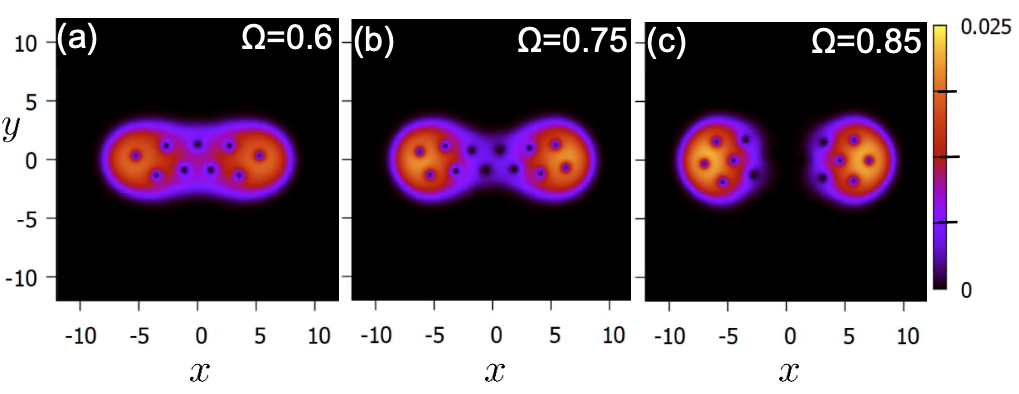}  
    \caption{
    For $r-$quartic potential [Eq~\eqref{eq16}], with $g=500$, $a=0.5$ and $b^2=0.02$ ($gb^4=0.2$), these three panels, 
    for $\Omega=0.6$ (a), 0.75 (b) and 0.85 (c), are exemplifying the density profiles corresponding,
    respectively, to low frequency conditions, before, close to, and after the breakup. With the defined units, all quantities 
    are dimensionless. }
    \label{fig07}
\end{figure}

\begin{figure}[!htbp]
    \centering
\includegraphics[scale=.55]{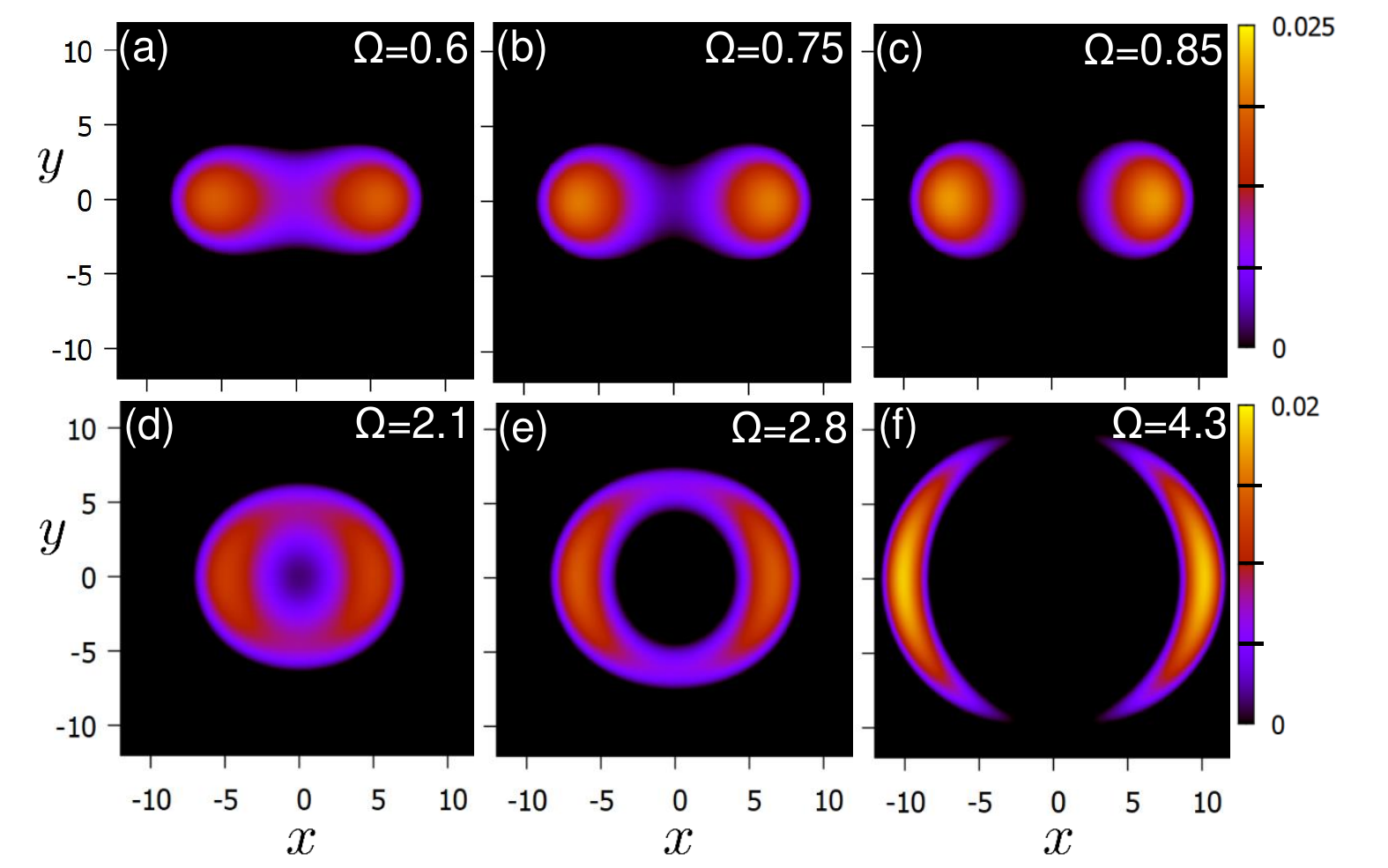}   
    \caption{Densities by using the TF approximation, for the $r-$quartic potential. The upper panels are for low frequencies, 
   as in Fig.~\ref{fig07}, with $g=500$, $a=0.5$, $b^2=0.02$ ($gb^4=0.2$). 
    The lower panels are for high frequencies, when giant vortex and breakup can be verified, with $g=2500$, $a=0$, $b^2=0.18$
    ($gb^4=81$).  The values of $\Omega$ are indicated inside the panels.  With the defined units, all quantities are dimensionless.}
    \label{fig08}
\end{figure}

With Fig.~\ref{fig06} we resume our results for the critical frequencies in terms of $gb^4$, considering two values
for the parameter $a$ (=0 and 0.5). 
For a better display of the lower rotation frequency, $\Omega<1$, the results are presented in log-scale in the inset. 
For such cases with $\Omega<1$, we can only observe breakup of the condensate, with the limiting
condition $\Omega_c$ given when the density at the center reduces to zero.  To illustrate this case, we have the density 
distribution of the condensate shown in Fig.~\ref{fig07}, where we include three panels with sample results, obtained by
the numerical GP results.
With $\Omega=0.6$ [panel (a), before the breakup], $\Omega=0.75$
[panel (b), close to the breakup], and $\Omega=0.85$ [panel (c), after the breakup].
 To further appreciate the utility of the TF approximation for the analysis of  the radial-quartic trap confinement, we 
 are also including the Fig.~\ref{fig08} with several panels, in which the upper panels [(a), (b) and (c)], with $\Omega^2<1$,  
 are in agreement with the same numerical results already presented in Fig.~\ref{fig07} (parameters $a=0.5$, $b^2=0.02$ and
 $g=500$).  As also indicated in Fig.~\ref{fig06}, no giant vortex can be formed for these low-rotation frequency cases, but 
 just breakup is possible. In the lower panels of Fig.~\ref{fig08}, we have also TF results with $\Omega^2>1$, corresponding 
 to potential parameters given by $a=0$, $b^2=0.18$ and g=2500, for $\Omega=$2.1 [(d), just below the formation of a
 giant vortex], 2.8 [(e), after the giant vortex is formed] and 4.3 [(f), after the breakup].  All these cases are indicated in
 the Fig.~\ref{fig06}; with $gb^4$ given in a log-scale inset panel for lower frequencies, and with $gb^4$ in normal scale 
 shown in the main panel for higher frequencies.


\begin{figure}[!htbp]
    \centering
    \includegraphics[scale=1.1]{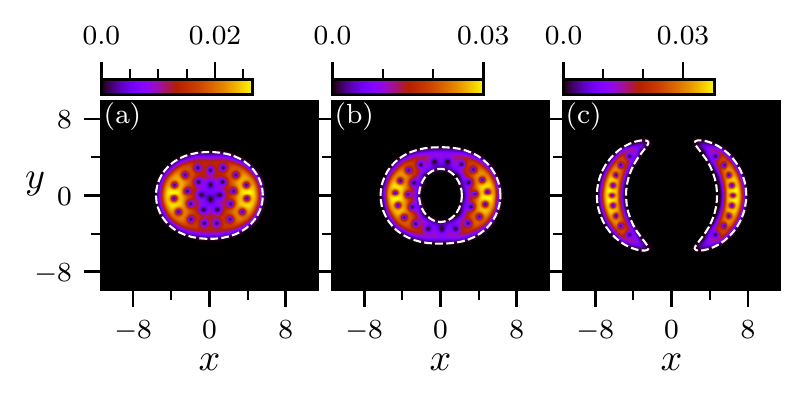}
    \includegraphics[scale=0.12]{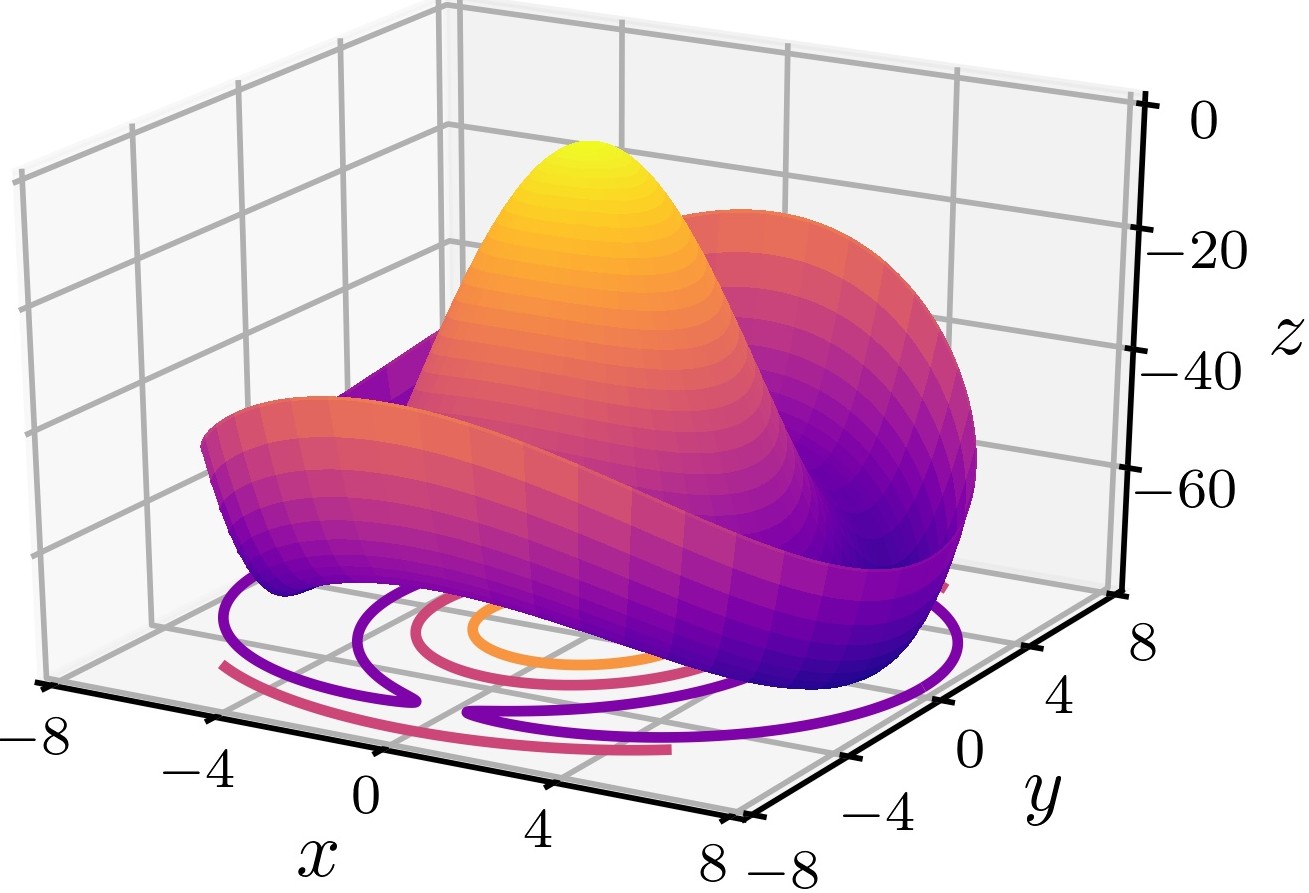}
    \caption{The upper panels show $|\psi|^2$, by using the Eq.~\eqref{eq16}, with $g=500$, 
    $b^2 = 0.18$  ($gb^4=16.2$) and $a = 0$, considering $\Omega = 1.65$ (a), 2.00 (b) and 2.75 (c). 
    These panels are exemplifying specific points of Fig.~\ref{fig06}, with giant hole in (b) and after the breakup, in (c).
    The TF contour are shown by the white-dashed lines, obtained as solution of Eq.~\eqref{eq21}.
    The surface plot shows the effective potential corresponding to panel (c) [$z\equiv V_1(x,y)-(2.75)^2/2$]. 
    With the defined units, all quantities are dimensionless. }
    \label{fig09}
\end{figure}

\begin{figure}[!htbp]
\centering
\includegraphics[scale=.26]{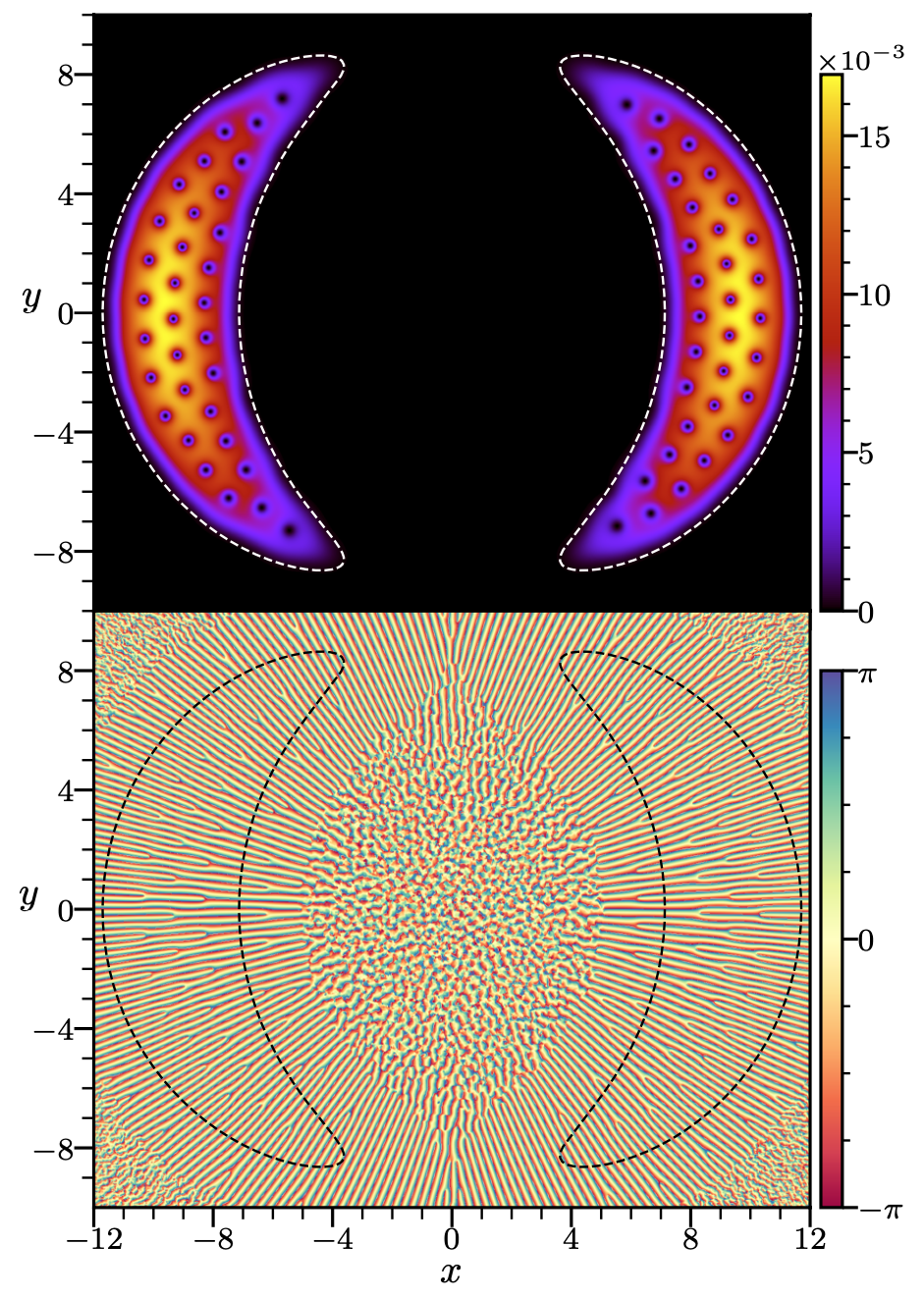}
\caption{
The density with vortex distributions (upper panel), together with the corresponding phase diagram (lower panel), 
are shown for $g=2500$, with $a=0$, $gb^4=16.2$ and $\Omega=2.75$, as in Fig.~\ref{fig09}(c) [implying $b^2\approx 0.08$]. 
The results are from numerical GP equation, with the dashed contour lines given by the TF approximation.  
With the defined units, all quantities are dimensionless. 
}    \label{fig10}
\end{figure}

When considering higher frequencies,  $\Omega>1$, as we increase $\Omega$ (for given potential parameter), we first observe a 
giant vortex being created at the center, for which the limiting $\Omega$ is identified by $\Omega_G$, represented in 
Fig.~\ref{fig06} by the dashed lines. In order to verify the splitting of the cloud, we need to increase $\Omega$, 
even more, such that $\Omega_G<\Omega$, with the limiting given by $\Omega_s$, represented in Fig.~\ref{fig06} 
by the solid lines.
This case with  $\Omega>1$ is also illustrated with three panels in Fig.~\ref{fig09}, for $\Omega=$1.65 [panel (a), before the
giant vortex is created], $\Omega=$ 2 [panel (b), where we observe the giant vortex at the center, before the breakup],
and $\Omega=$2.75 [panel (c), after the breakup]. In these three panels, we also show that the TF approximation can give
quite precisely the contour of the cloud, along all the stages of the rotation-frequency conditions; namely, 
before the creation of the giant vortex, during and after that, and even after the cloud is fragmented in two parts.
We are also illustrating this figure with a 3D surface representation of  the effective potential, corresponding to the 
panel (c).

 In Fig.~\ref{fig09}, the panels (a), (b) and (c) refer to the three points shown in 
Fig.~\ref{fig06} for $gb^4=16.2$.
One may clearly note the centered-hole phase in Fig.~\ref{fig09} (b) and the breakup in Fig.~\ref{fig09} (c).
Corresponding to the three specific cases shown in Fig.~\ref{fig09}, together with the $\Omega=0$ case,  
in Table~\ref{tab:Mu2} we have the TF and GP chemical potentials, with the GP total energy.
\begin{table}[ht]
\caption{
The chemical potentials (GP numerical results and TF approximation), for $\Omega=0$ and for the three cases presented
in Fig.~\ref{fig09}, are shown in this table.
The units are $\hbar\omega$ for the energy and chemical potentials,  and $\omega$ for $\Omega$.}
\label{tab:Mu2}
\centering
    \begin{tabular}{ c c c c }
    \hline\hline
$\Omega$&$\mu$        &$\mu_{TF}$ &   Energy      \\ \hline \hline
        0&15.787&15.637& 9.814\\
        1.65  & 3.808  &  1.484  &  -0.199  \\
        2  & -6.082  &  -8.922  &  -10.274  \\
        2.75  & -52.490  &  -62.741  &  -57.417 \\
\hline\hline
    \end{tabular}
\end{table}

In view that it was clearly shown that the TF approximation works quite well to define the contour of the condensate, 
as shown in the three panels of Fig.~\ref{fig09}, we found useful also to verify the effect of increasing the density factor 
from $g=500$ to $g=2500$, by considering the panel (c) of  Fig.~\ref{fig09}. For that, in Fig.~\ref{fig10} we kept the
same values $gb^4$ used in Fig.~\ref{fig09}, with $a=0$ and $\Omega=2.75$. As shown, a length-scale transformation is 
observed, which is also reflected in the corresponding density.
The length-scale transformation is transparent in the Eqs.~\eqref{eq17}, \eqref{eq18} and \eqref{eq19}. As one can
see clearly, we have a scale transformation, such that $(\tilde{x},\tilde{y})=(bx,by)$ implies 
$gn(x,y)\to [gb^2 \tilde{n}(\tilde{x},\tilde{y})]$ (with $\int d\tilde{x}d\tilde{y}\,
\tilde{n}(\tilde{x},\tilde{y})=1$).

By considering this scaling transformation, it is possible to be verified that both results, shown in Fig.~\ref{fig09}(c) and 
Fig.~\ref{fig10}, will correspond to the scaling of all the corresponding physical quantities, as the trap interaction, 
chemical potential and energies (which will be multiplied by a factor $b^2$).  Within this scaling procedure, 
the strength of the nonlinear term will be rescaled, such that $g\to\overline{g}=gb^4$.
Therefore, the net effect due to the kinetic energy, which is obtained by solving the full
GP formalism, will be reflected on the relation of this energy with the frequency parameter $\Omega$,
with the increasing vorticity of the system.  
 
 Another aspect which may be of interest, is to estimate numerically the free expansion of the condensate, 
by switching off the trap, in order to check with possible experimental realizations.
By considering a limited short time interval, with $t<16$ (units $\omega^{-1})$, we have verified that the vortex patterns 
and corresponding numbers are kept about the same during the expansion, when considering a case with high frequency
in which a giant vortex can be observed  at the center. As expected, the density starts being reduced as the condensed 
cloud expands, with the size of the giant vortex increasing at the same proportion.
However, we understand that the required time period for numerical simulations should be much larger,
covering an increasing spatial distribution, in order to draw a more conclusive picture about this dynamics, such that could
be more useful for experimental studies. This is a perspective theoretical investigation of great interest to be performed 
within a full dynamical study in which one could introduce time-dependent parameters in the trap interaction. 

\section{Vortex pattern distribution and the Feynman rule}\label{sec3}
As observed from the numerical results, the vortex pattern distribution follows approximately the
corresponding space occupied by the condensate cloud. The number of vortex is related to the angular frequency of 
the rotation and the space occupied by a rotating superfluid, which is the well-known Feynman rule (FR)~\cite{1955-Feynman}
established when studying the liquid helium (For a detailed analysis, considering rapidly rotating condensate with 
dense array of vortices, see Ref.~\cite{2009Fetter}).
According to this rule, the number of vortices $N_v$ inside a given area $A$ is linearly proportional to the rotation 
frequency $\Omega$, which in our dimensionless units can be written as $N_v= \Omega  A/\pi$. 
As pointed out in \cite{2010-Wen}, this rule was initially formulated for uniform superfluid helium, been intensively studied
both theoretically and experimentally for a BEC trapped in a single harmonic potential. 
So, it is of interest to verify the extension of the rule when considering trap geometries other than symmetric ones. 
In our case, by considering the results shown in Fig.~\ref{fig02}, we noticed that apparently it fails if we consider the 
visible vortices within the full area defined by the TF boundaries.
However,  one should realize that may not be the correct way to apply the rule, in particular when the area is not regular,
having low-density regions, such that the visible vortex distribution becomes not uniform. 
In this regards, the discrepancy in the applicability of the FR has been observed before for the case of double-well trap 
confinement~\cite{2010-Wen}, being attributed to hidden vortices. 

In order to clarify that, let us first consider the cases represented in Fig.~\ref{fig03}, where the nonlinear term is much larger 
than in Fig.~\ref{fig02}, with $g=2500$. Notice that, after the breakup, when $\Omega=0.77$ [panel (b)], the densities 
of the two fragments are more or less established visibly within two triangle formats, if one considers the limiting space
defined by the TF contours. In this case,  we can visually approximate the total area as the corresponding one occupied 
by two triangles, having $A=2\times 90$ (units of $l_\omega^2$). With that,  by applying the FR just for the visible vortices, 
we obtain  $N_v\approx 0.77\times (180/\pi)\approx 44$, which matches exactly with the observed visible vortices inside 
the area defined by the TF boundary. In the other case, before the breakup,
shown by the panel (a) of Fig.~\ref{fig03}, with $\Omega=0.7$, we have 41 visible vortices inside the TF contour limits, 
also in accordance with the corresponding FR prediction. 
So, the assumption presented in Ref.~\cite{2010-Wen}, as requiring hidden vortices to enter in the estimated value of
$N_v$,  apparently fails when considering these specific results. Let us try to clarify that. 
By considering the respective phase diagrams for the panels (a) and (b), we can clearly verify the existence
of hidden vortices in the low-density regions (see, in particular near the $x=0$ positions). However, they are outside 
the area defined by the TF contour, where we are applying the Feynman rule. More precisely, they are emerging only 
when we consider a more extended area for the densities, which is given by the GP numerical solution.
So, once consistently we define the respective area in which enough vortices exist, the applicability of the FR is 
observed. Suppose, for instance, that we ignore the TF limiting space, considering a circular area with center at $(x,y)=0$ 
and arbitrary radius, covering a space such that we include the hidden vortices near $x=0$. 
Again we could verify, approximately, the applicability of the Feynman rule; this time, including the hidden vortices. 

In resume, the possible regularity of the trap can only be used to explain visual discrepancy in the applicability of the
FR if, inside the given specific region there are hidden vortices, which need to be included. 
In Figs.~\ref{fig02} and \ref{fig03} we have about the same confining potential shape.  So, for small non-linearity, as with 
$g=500$, the rule apparently is requiring hidden vortices to be verified (present inside the TF limiting contour),
 but not in the second case.  
In conclusion, the visual verification of the FR in a large area requires an enough dense cloud, which can support 
larger number of vortices, as in the second case, when we have
the nonlinear parameter $g=2500$ (indicating larger condensed cloud). With enough dense cloud, one can select
an arbitrary specific area inside the condensate to apply the rule.

By considering a minimum area $A=\pi R_m^2$ to have one vortex inside, the rule can be expressed by 
\begin{equation}
\eta_v\equiv \frac{N_v}{A} = \frac{\Omega}{\pi},\;\;\to\;\; {R_m}=\sqrt{\frac{1}{\Omega}},
\label{eq29}\end{equation} 
which also implies in about $2R_m$ the distance between two vortices.
From this reading, the FR can be approximately verified, not only for the case that we have a condensate enough 
dense with $g=2500$ (where we have verified complete agreement), but even for the case that we have $g=500$, 
shown in Fig.~\ref{fig02}. This is shown by the Table~\ref{tab-FR}, where $R_m$ comes from Eq.~(\ref{eq29}),
with $D$ obtained visually from the respective panels. 
In this way, the applicability of the FR can also be extended by including low-density regions, where regular 
patterns of hidden vortices can be detected, by examining the corresponding phase diagrams. See, for example,
the results shown in the Fig.~\ref{fig10}, in which hidden vortices (not visible in the upper panel)
can be identified (outside the TF contour line) in the corresponding phase diagram.
\begin{table}[ht]
\caption{Applicability of the Feynman rule in the given examples of this contribution, Figs.~\ref{fig02}-\ref{fig10}.
$2R_{m}$ is the predicted distance between two vortices. $D$ is a visual approximate distance between 
the closest vortices (extracted near the minima of the effective potential, with possible $\pm$0.04 error).
}
\label{tab-FR}
\centering
    \begin{tabular}{ c |c c c c c }
    \hline\hline
Fig.&$Panels$    & $\Omega$& $2R_{m}$ & $D$ & $g$ \\ \hline \hline
\ref{fig02}&(a)&        0.3    & 3.65 &
(\footnote{Not applicable, as we have only one vortex near each minimum.})&500   \\
&(b)&        0.4    & 3.16 & 3.00 &500   \\
&(c)&        0.5    & 2.83 & 2.60 &500  \\
&(d)&        0.6    & 2.58 & 2.40 &500   \\ \hline
\ref{fig03}
&(a)&        0.70   & 2.39& 2.40 &2500  \\ 
&(b)&        0.77   & 2.28 & 2.30 &2500  \\
\hline
\ref{fig07}&(a)&        0.60   &2.58&2.50 &500 \\
&(b)&        0.75   &2.31&2.25 &500 \\
&(c)&        0.85   &2.17&2.00 &500 \\ \hline
\ref{fig09}&(a)&        1.65   &1.56&1.46 &500 \\
&(b)&        2.00   &1.41&1.37 &500 \\
&(c)&        2.75   &1.21&1.09 &500 \\ \hline
\ref{fig10}&-&        2.75   &1.21&1.20&2500 \\
\hline\hline
    \end{tabular}
\end{table}
\section{Conclusions}\label{sec4}
The parameter conditions for a single ultra-cold rotating Bose-Einstein condensate to be fragmented, as well as 
to form giant vortex at the center, were investigated in this work, by using two kinds of asymmetric pancakelike 
external quartic-quadratic potentials.  The corresponding GP formalism was solved by numerical approach, as well as 
by a detailed analysis using the TF approximation. The computational results were obtained for low and high 
rotational frequencies, describing the usual vortex-pattern distributions. In this case, the vortex density distributions 
was found to follow closely the well-known Feynman rule, once we have appropriately defined the specific region 
where enough vortices can be observed.  

Of particular relevance, considering that it can be quite useful in experimental setup analysis, is TF approximation 
reliability to define the limiting frequencies for the density distribution of the condensate, even when considering
the quite non-uniform condensate distributions.  The space contour limits of the condensate, as well as the
averaged density distributions, are shown to be well described by the TF approximation, as compared with the 
GP numerical results, even after the breakup of the condensate, which happens as the rotation frequency increases. 
These results, verified for different quartic-quadratic confinement, and also by varying the repulsive interaction
parameter, provides some further support to the TF approximation even when considering single atomic systems 
with more complex properties.

As expected, the vortex pattern picture is provided by the GP numerical approach, in which we noticed a 
quite regular triangular shape structure for the vortices, when the rotation is enough high, before the 
condensate breakup. The vortex distributions, close to their symmetrically distributed maxima (minima of the effective
potential),  apparently are not strongly affected by the breakup conditions when higher rotation are implemented.
In these cases, we have also verified that the visible vortex pattern distributions follow the well-known 
Feynman rule, when applying this rule for sample spacial region with enough number of vortices. With 
this rule given in terms of the vortex density, we can define a unit of area in order to apply the FR even to 
low-density regions, where hidden vortices are more likely to appear.

The present study for the breakup conditions in rotating condensate was done for the simplest possible situation 
in which we consider an asymmetric quartic-quadratic trap confinement, enough large to control possible higher 
rotation conditions, in which the condensate fragments are still kept confined.
The possible relevance of the present investigation relies in the fact that we are providing quite well defined 
limiting conditions via TF approximation (to our acknowledgment, not presented before in the usual literature),
which can be easily followed in actual experimental realizations,
with different condensed atomic samples, trapped in quartic-quadratic pancakelike potentials. 

We understand this kind of study can be quite useful particularly when designing experimental setups of other more
involved investigations, such as with dipolar single-component BEC systems, binary miscible and immiscible combinations, etc.
As a perspective interesting study, one could further investigate the confined system by assuming time-dependent rotation 
frequency or trap configurations, which could be introduced by using the present approach. By manipulating the trap configuration
as shown, for example, in this work, one could implement another mechanism to control the condensed cloud, in addition to 
the use of Feshbach resonance techniques.

\noindent{\bf Acknowledgments}\\
We are grateful to Romain Dubessy for his interest and helpful discussion.
The authors thank the Brazilian agencies Funda\c{c}\~ao de Amparo \`a Pesquisa do Estado de S\~ao Paulo
(FAPESP)  [Contracts 2018/02737-4(AA), 2017/05660-0(LT), 2016/17612-7 (AG)], 
 Conselho Nacional de Desenvolvimento Cient\'\i fico e Tecnol\'ogico [Procs. 304469-2019-0(LT) and 
306920/2018-2 (AG)] and 
Coordena\c c\~ao de Aperfei\c coamento de Pessoal de N\'\i vel Superior [Proc. 88887.374855/2019-00 (LB)].

\end{document}